\documentclass[twocolumns,traditabstract]{aa}  

\usepackage{changepage}
\usepackage{tabularx,multirow,booktabs,blindtext}
\usepackage{adjustbox}
\usepackage{graphicx}
\usepackage{subfig}
\usepackage{placeins} 
\usepackage{txfonts}
\usepackage{graphicx,rotating}
\usepackage{subfig}
\usepackage{txfonts}
\usepackage[T1]{fontenc}
\usepackage{epsfig}
\usepackage{natbib}
\usepackage{color}
\usepackage{amssymb}  
\usepackage{tabularx}
\usepackage{comment}
\usepackage{longtable}
\usepackage{ltablex}
\usepackage{caption, booktabs}
\def\mgii{Mg\,{\sc ii}}
\def\civ{C\,{\sc iv}}
\def\halpha{H$\alpha$}
\def\hbeta{H$\beta$}
\DeclareGraphicsExtensions{.pdf,.png,.jpg}
\bibpunct{(}{)}{;}{a}{}{,} % to follow the A&A style
\begin{document}

\author{Rémi Poitevineau} % Your name

%\date{\normalsize\today} % Today's date or a custom date

% \begin{document}

   \title{Black hole and galaxy co-evolution in radio-loud active galactic nuclei at $z\sim0.3-4$}

   \author{
    R. Poitevineau \inst{1}\fnmsep\thanks{e-mail: remi.poitevineau@obspm.fr}  
	 \and  
     G. Castignani \inst{2,3}
     \and
      F. Combes\inst{1,4}
     %\and
     %...
%      C. Benoist\inst{4}
      }

   \institute{Observatoire de Paris, LERMA, PSL University, Sorbonne Universit\'{e}, CNRS, F-75014, Paris, France  
            \and
            Dipartimento di Fisica e Astronomia ''Augusto Righi'', Alma Mater Studiorum Università di Bologna, Via Gobetti 93/2, I-40129 Bologna, Italy   
            \and
            INAF - Osservatorio  di  Astrofisica  e  Scienza  dello  Spazio di  Bologna,  via  Gobetti  93/3,  I-40129,  Bologna,  Italy
            \and
            Coll\`{e}ge de France, 11 Place Marcelin Berthelot, 75231 Paris, France
            %\and
            %Universit\'{e} C\^{o}te d'Azur, OCA, CNRS, Lagrange, UMR 7293, CS 34229, 06304, Nice Cedex 4, France
        }

                \date{Received 21 July 2022}
                
% REPORT RECEIVED ON JAN 12 2023
% \abstract{}{}{}{}{} 
% 5 {} token are mandatory
 
  \abstract
  {The relation between the mass of the supermassive black hole (SMBH) in the center of galaxies and their bulge mass or central velocity dispersion is well known. This suggests a coevolution between the SMBHs and their galaxy hosts. Our aim is to study this relation, specifically, for radio loud galaxies, and as a function of redshift $z$. We selected a sample of 42 radio galaxies and active galactic nuclei (AGN) with broad emission lines and  spectroscopic redshifts between $z=0.3-4$ by cross-matching the low radio frequency sources from Very Large Array (VLA) FIRST with spectroscopically confirmed galaxies from wide-field surveys, including Sloan Digital Sky Survey (SDSS) DR14 \textsf{ugriz} and  Dark Energy Survey (DES) DR2 \textsf{grzY} in the optical, Wield Infrared Survey Explorer (WISE), and the Galaxy And Mass Assembly (GAMA) spectroscopic survey.  
  We characterized the stellar mass ($M_{\star}$), star formation, and black hole properties (mass of the central SMBH, Eddington ratio $\eta$ , and jet power, $Q_{\rm jet}$). The relation between SMBH mass, $M_\star$, $\eta$ , and $z$ is placed into context by comparing them with scaling relations ($M_{\rm BH}$--$M_{\star}$, $M_{\rm BH}/M_\star$--$z$,  $M_{\rm BH}$--$Q_{\rm jet}$ , and $Q_{\rm jet}$--$\eta$) from the literature. 
  On the basis of a multiwavelength spectral energy distribution modeling, our radio sources are broadly consistent with being on the star-forming main sequence. They have sub-Eddington accretion rates, $\eta\simeq1\%$ on average, as typically found in type I AGN, while higher accretion rates favor more powerful jets to be launched by the central engine. We find overmassive SMBHs in $(17\pm5)\%$ of our radio sources, similarly to previous studies on nearby early-type galaxies. Altogether, an evolutionary scenario in which radio-mode AGN feedback regulates the accretion onto the SMBHs and the stellar mass assembly of the radio sources is discussed, which may explain the observed phenomenology. This pilot study represents a benchmark for future studies using wide-field surveys such as those with Euclid and the Vera Rubin Observatory.}


   \maketitle
%
%-------------------------------------------------------------------

\section{Introduction}

Supermassive black holes (SMBHs), characterized by masses in the range $\sim 10^{6}~M_{\odot}$ to $\sim 10^{10}~M_{\odot}$, are observed to lie at the center of most, if not all, massive galaxies \citep[e.g.,][]{graham_galaxy_2016}.
When the central regions of galaxies are sources of radiation because
of accretion onto their SMBHs, they are called
active galactic nuclei (AGN). AGN are among the strongest proofs for the existence of SMBHs,
together with the direct measure of high densities in our Galactic center \citep[e.g.,][]{genzel_galactic_2010},
and the direct observation of the SMBH shadow at the center of M87  and of the Milky Way itself \citep{EHT_M87_2019,EHT_MW_2022}.

An intrinsic coevolution exists between AGN activity, SMBH growth, galaxy stellar content, and star formation history \citep[e.g.,][for a review]{kormendy_coevolution_2013}. In some cases, AGN are jetted and are then called radio-loud AGN.  They constitute only 10\% of the whole AGN population, but their fraction varies with the stellar mass of the host, from 0 to 30\% \citep{best_host_2005}.
Large-scale radio jets are even able to impact the global megaparsec-scale environmental properties via radio-mode AGN feedback, for example, at the center of galaxy (proto-)clusters \citep[e.g.,][for a review]{Miley_DeBreuck2008,fabian_observational_2012,Magliocchetti2022}.

The mode of SMBH accretion ultimately regulates the excitation properties of radio-loud AGN. It is indeed possible to distinguish two main classes of activity among the radio-loud AGN: high-excitation (HE) and low-excitation (LE) radio galaxies (RG) according to their accretion rate: HERGs typically have accretion rates between 1 and 10\% of their Eddington rate, whereas LERGs predominately accrete at a rate below 1\% Eddington \citep{best_fundamental_2012}. In HERGs, the material thus progressively loses angular momentum in a geometrically thin disk around the SMBH; this disk is usually optically thick and radiates efficiently. When the accretion rate is below 0.01 the Eddington rate, the AGN is instead characterized by an advection-dominated accretion flow \citep[ADAF; e.g.,][]{narayan_advection-dominated_2008}, which radiates inefficiently. Radio-loud AGN mostly occur in the low-luminosity regime, and ADAFs frequently occur as well.

A major observational breakthrough for the coevolution of galaxies and AGN with their SMBHs  was the discovery of a tight correlation in the local universe between the SMBH mass and the mass of their host spheroids \citep[][]{magorrian_demography_1998,Ferrarese_Merritt2000}.  This relation implies a remarkable connection between the assembly of galaxies and the formation and growth of SMBHs at their center \citep[e.g.,][for a review]{Heckman_Best2014}. 
Models and simulations \citep{Menci2006, Marulli2008, Hopkins2006,Volonteri_Natarajan2009}
have attempted to explain this correlation and its evolution with redshift, as found in several observational studies \citep[e.g.,][]{shields_black_2006,sarria2010,wang_molecular_2010,merloni_cosmic_2009,jahnke_massive_2009,cisternas_secular_2011,schramm_black_2013}.

There is, however, still a number of related open issues. These include local ellipticals with overmassive SMBHs \citep{Kormendy1997,van_den_Bosch2012,Savorgnan_Graham2016,Dullo2021}. These overmassive SMBH preferentially occur in galaxy clusters and in brightest cluster galaxies in particular \cite[BCGs; e.g.,][]{mcconnell_revisiting_2013}, where environment effects strip the galaxies from their gas and stop star formation and the growth of bulges. Galaxies are then called massive relics and have particularly old stellar population \citep{Trujillo2014, Martin-Navarro2015, Ferre-Mateu2015, Ferre-Mateu2017}.
The very discovery of massive SMBHs ($M_{\rm BH}\gtrsim 10^{9}~M_{\odot}$) in bright quasars at the epoch of reionization \citep[e.g.,][]{banados_800-million-solar-mass_2018,Farina2022} is a mystery, as it shows that extreme SMBHs can form within 1 Gyr after the Big Bang. The rapid formation of these high-$z$ SMBHs might be explained by invoking some extreme scenarios such as the growth of a $10^{2-5}~M_{\odot}$ seed via super-Eddington accretion \citep{valiante_first_2016,pezzulli_sustainable_2017}, the direct collapse of an initial gas condensation into a black hole of $\sim$10$^5$ M$_\odot$
\citep{visbal_direct_2014, regan_rapid_2017}, or the merger of massive protogalaxies
\citep[e.g.,][]{mayer_direct_2010, mayer_direct_2015, ferrara_can_2013, bonoli_massive_2014}.

Altogether, while existing studies show a tight coevolution of SMBHs, AGN, and their host galaxies with cosmic time, this interplay is still substantially debated and unconstrained. This is at least partially due to the difficulty of building large samples of distant AGN with well-characterized stellar and black hole properties.

In order to better understand the growth of SMBHs with cosmic time and their coevolution with their host galaxies, we have built a sample of distant radio-loud AGN spanning about 9~Gyr of cosmic time, between $z\sim0.3-4$, with available radio to ultraviolet (UV) spectrophotometric data. Based on this multiwavelength dataset, we assess the properties of the AGN sample, for example, in terms of black hole and stellar masses, jet power, and Eddington ratio. As radio-loud AGN are associated with the most massive black holes and host galaxies \citep[e.g.,][]{best_host_2005,Chiaberge_Marconi2011,shen_catalog_2011,shaw_spectroscopy_2012}, they are excellent sources in which to investigate the galaxy, AGN, and SMBH coevolution in the high-mass regime.

The paper is organized as follows. In Sect.~\ref{sec:sample} we describe our sample selection as well as its multiwavelength dataset and properties.  In Sect.~\ref{sec:BH_jet_Mstar} we report estimates for the black hole, jet, accretion, and stellar properties.
In Sect.~\ref{sec:control} we describe our comparison sample.
The results in terms of the scaling relations of black hole, jet, and host galaxy are reported in Sect.~\ref{sec:results}.
In Sect.~\ref{sec:conclusions} we summarize the results and draw our conclusions.

Throughout this work, we adopt a flat $\Lambda \rm CDM$ cosmology with matter density $\Omega_{\rm m} = 0.30$, dark energy density $\Omega_{\Lambda} = 0.70,$ and Hubble constant $h=H_0/100\, \rm km\,s^{-1}\,Mpc^{-1} = 0.70$.

\section{Radio-loud AGN sample}
\label{sec:sample}
We selected a sample of radio-loud AGN by cross-matching the  Very Large Array Faint Images of the Radio Sky at Twenty-centimeters (VLA FIRST) source catalog \citep{becker_first_1995} with infrared (IR) to optical spectrophotometric surveys. As further described in the following, the use of IR to UV photometry enables modeling the spectral energy distribution (SED), which ultimately allows us to obtain a good characterization of the galaxy properties, such as the stellar mass ($M_{\star}$) and the star formation rate (SFR).

\subsection{{Dark Energy Survey}}\label{sec:DES}
{We started by considering the Dark Energy Survey \citep[DES][]{the_dark_energy_survey_collaboration_dark_2005,dark_energy_survey_collaboration_dark_2016}, which is} composed of two distinct multiband imaging surveys: a $\sim$5000~deg$^2$ wide-area \textsf{grizY} survey, and a deep supernova \textsf{griz} survey consisting of six distinct deep fields \citep{hartley_dark_2021}. The coadded source catalog and images from the processing of all six years of DES wide-area survey observations and all five years of DES supernova survey observations have recently been made public with the DES data release 2 \citep{abbott_dark_2021}\footnote{https://des.ncsa.illinois.edu/releases/dr2}.

{To build our sample of distant radio-loud AGN, we limited ourselves to equatorial DES supernova fields that overlapped with the VLA FIRST survey at low radio frequencies. The selection was similar to that of our previous work \citep[][]{castignani_molecular_2019}, to which we refer for further details. However, in that study,  we focused only on two radio sources, and we investigated their molecular gas content, their cluster environment, as well as the stellar and star formation properties. In this work, we consider instead a more extended sample, as further outlined in the following.}

\subsection{Radio, optical, and spectroscopic selection}
As we are interested in building a sample of extragalactic radio sources, we considered the VLA FIRST survey \citep{becker_first_1995}, which observed 10,000~deg$^2$ of the North and South Galactic Caps at low radio frequencies (1.4~GHz), down to a point source detection limit of $\sim$1~mJy.
We therefore further limited ourselves to the Stripe 82 area, that is, a 300 deg$^2$ equatorial field that was imaged multiple times by the Sloan Digital Sky Survey (SDSS) and overlaps with the VLA FIRST survey. Similarly, we {additionally} considered DES supernova deep fields numbered 2, 3, and 5 for our search, as outlined in Sect.~\ref{sec:DES}

We cross-matched the low radio frequency VLA FIRST radio source catalog with both the SDSS DR14 \textsf{ugriz} and DES DR1 \textsf{grizY} source catalogs within the considered fields with a search radius of 3~arcsec, consistent with the positional accuracy $\sim 1$ arcsec of FIRST sources. 
As we are interested in secure distant radio sources, we further restricted ourselves to sources with SDSS DR14 spectroscopic redshifts $z>0.3$.
The search yielded 158 spectroscopically confirmed radio sources with unique optical counterparts from both SDSS and DES.

\subsection{IR selection: WISE}\label{sec:WISEselection}
We further searched for IR emission of the radio sources, as found by the W4 filter of the Wide-field Infrared Survey Explorer  \citep[WISE;][]{wright_wide-field_2010}.
We therefore cross-correlated our radio sources with the allWISE source catalog\footnote{http://wise2.ipac.caltech.edu/docs/release/allwise/} by adopting a search radius of 6.5~arcsec, consistent with previous work on extragalactic radio sources \citep[e.g.,][]{castignani_black-hole_2013}. The search yielded 154 sources with unique WISE counterparts and W4 magnitudes with a signal-to-noise ratio S/N $>1$.

\subsection{Broad emission lines from SDSS}
As we are interested in assessing black hole masses of the considered radio-loud AGN, we further restricted ourselves to sources with evidence of broad emission lines in the SDSS spectra.  To do this, we further selected sources that have \halpha, \hbeta, \mgii, or \civ~ emission line fluxes at an S/N$>3$, as well as a full width at half maximum FWHM>1000~km~s$^{-1}$, typical of the broad-line region lines.  We used the spZline file\footnote{https://data.sdss.org/datamodel/files/BOSS\_SPECTRO\_REDUX/\\RUN2D/PLATE4/RUN1D/spZline.html} , which contains the results of the emission-line fits for the BOSS spectra of SDSS sources \citep[][]{bolton_spectral_2012}. The Gaussian line width $\sigma$ is  reported, and we converted it into the ${\rm FWHM}=2\sigma\sqrt{2\log 2}$. This additional spectroscopic selection yielded 21 sources at $z\sim0.3-3.8$.

\begin{figure}[hbt!]
\centering
\subfloat{\includegraphics[width=0.5\textwidth]{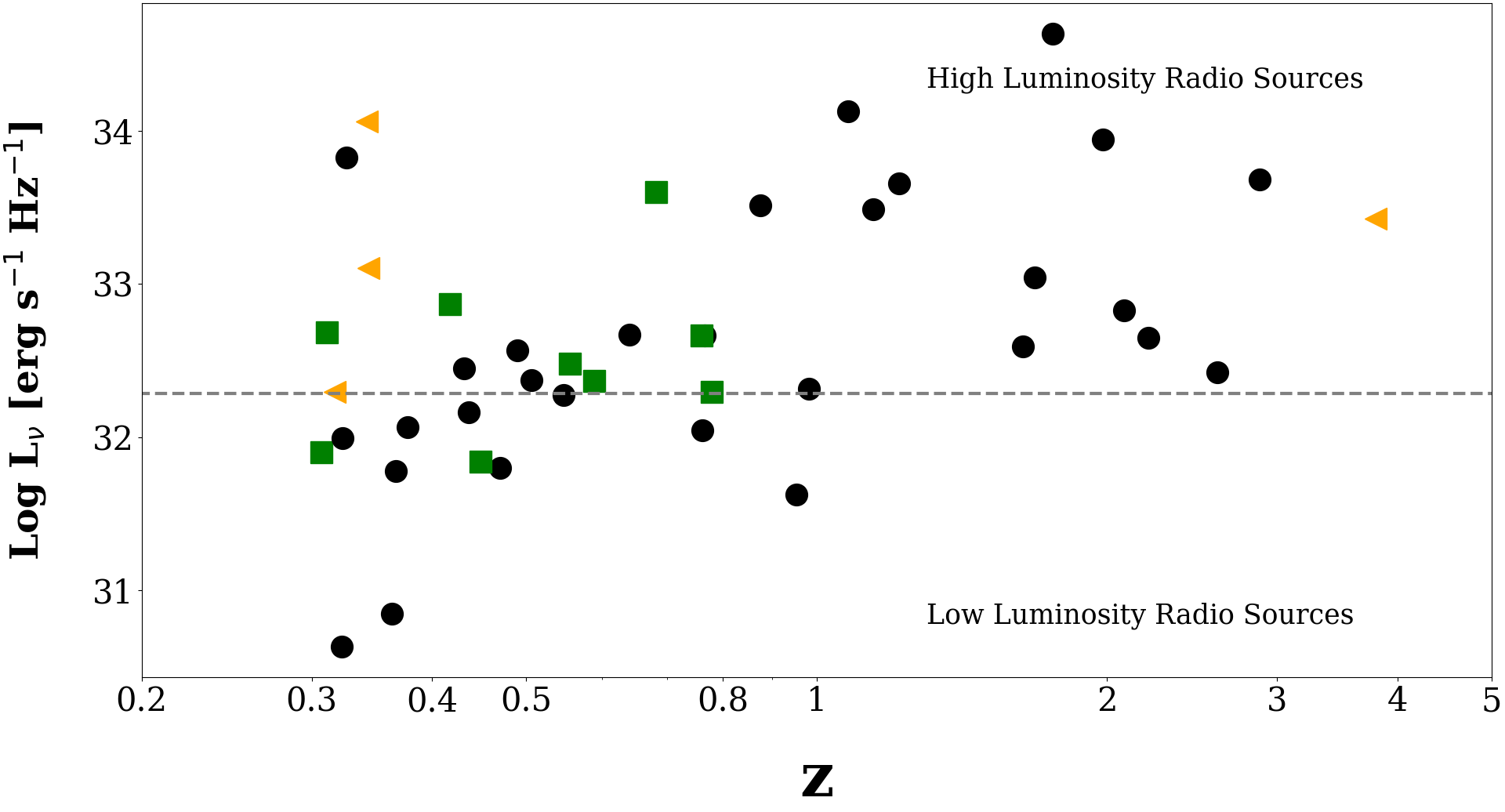}}\\
\subfloat{\includegraphics[width=0.5\textwidth]{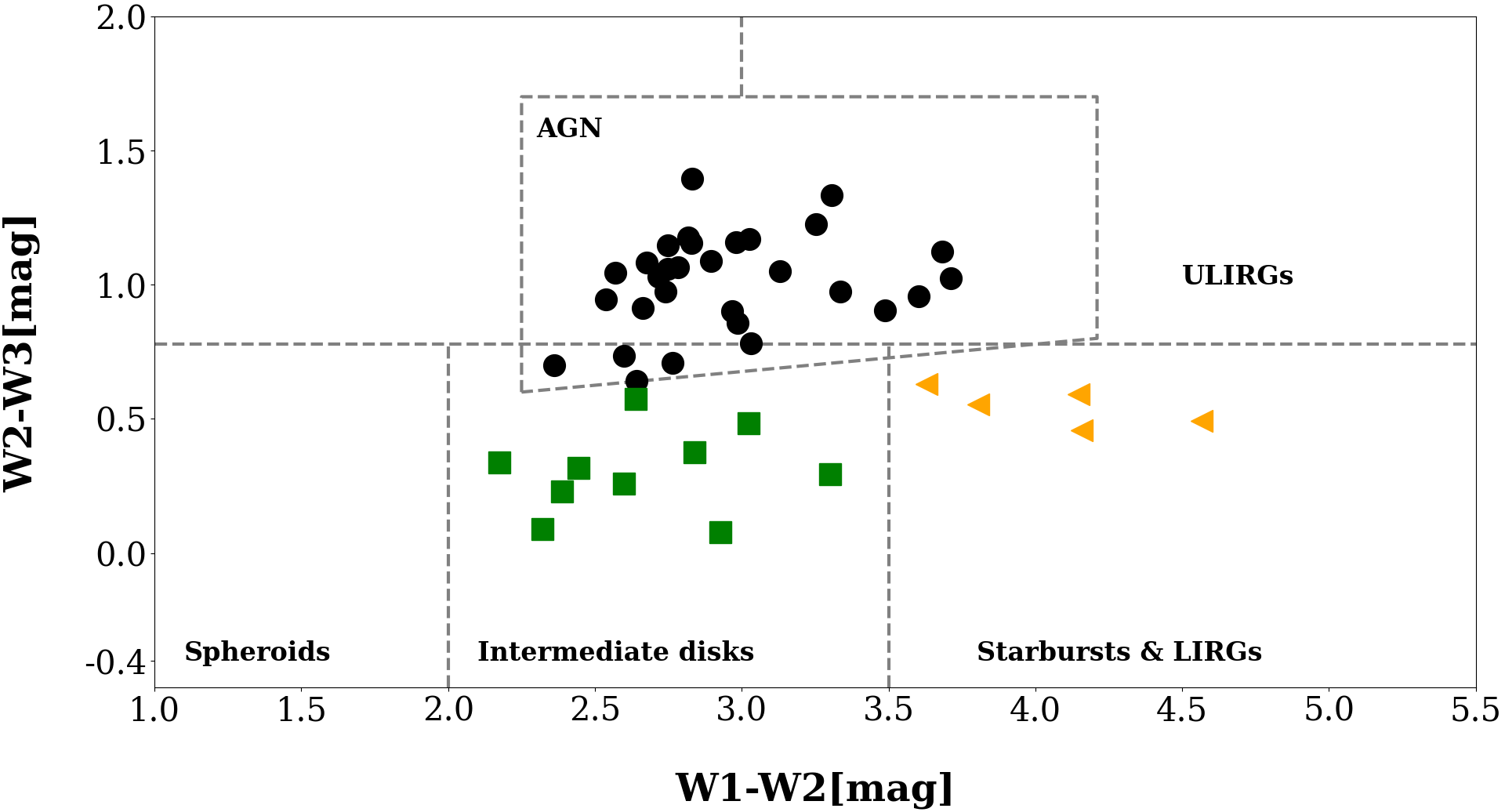}}
\caption{Radio luminosity densities and infrared colors for the radio sources in our sample. {\it Top.} 1.4~GHz Radio luminosity density as a function of redshift. Sources are distinguished between AGN (black circles), starbursts (orange triangles), intermediate disks (green squares), and spheroids according to the color-based classification by \citet{stern_mid-infrared_2012} and \citet{jarrett_galaxy_2017}. The horizontal line is at $L_\nu=2\times10^{32}~{\rm erg}~{\rm s}^{-1}~{\rm Hz}^{-1}$ and separates low-luminosity radio sources from high-luminosity radio sources.
%%%%%%%%%%
{\it Bottom.} WISE color-color plot. Color coding of the data-points is the same as in the top panel. Black dashed lines are overplotted to distinguish the different types of objects (AGN, ULIRGs, spheroids, intermediate disks, starbusts, and LIRGs).}
\label{fig:WISEcolors_radio_luminosity}
\end{figure}

\subsection{Radio-loud AGN in GAMA DR3}
{In addition to the radio-loud AGN selected as described in the previous subsections,} we searched for distant radio sources from the  third data release (DR) of the Galaxy And Mass Assembly (GAMA) spectroscopic survey. GAMA DR3 \citep{baldry_galaxy_2018} provides spectra obtained with the AAOmega multi-object spectrograph on the Anglo-Australian Telescope (AAT) as well as a wealth of ancillary information for more than 200 000 sources.

Similarly to what we did concerning the SDSS spectra, we first selected 632 sources at $z>0.3$, with available GAMA DR3 spectra, \halpha~ or \hbeta~ line fluxes at S/N$>3$ as well as line widths FWHM$>1000$~km~s$^{-1}$, as inferred from single Gaussian modeling. \footnote{http://www.gama-survey.org/dr3/}
We selected these sources within the equatorial area of 180~deg$^2$ covered by GAMA DR3 as well as by the 1.4~GHz FIRST VLA survey. For the 632 spectroscopic sources from GAMA, we further limited ourselves to the subsample of 39 galaxies with available 1.4~GHz fluxes from the FIRST VLA survey.
Multiple spectra are often available in the GAMA DR3 database. We then inspected each of the available spectra and discarded galaxies with dubious evidence of emission lines, which means that the associated fits are dubious as well. This analysis yielded 21 GAMA DR3 radio sources at $z\sim0.3-0.8$. We assigned unique WISE counterparts to them using a 6.5~arcsec search radius, as in Sect.~\ref{sec:WISEselection}.

By combining these galaxies, denoted hereafter with the prefix G, with the 21 radio sources with SDSS spectra, denoted with the prefix RS, our final sample comprises 42 sources that we consider hereafter for this study. The main properties of this sample of galaxies are listed in Tables~\ref{tab:A1sample}.

\subsection{Radio and IR properties}\label{sec:radio_IR_prop}
We now investigate the low-frequency radio luminosities and the IR colors of our sources. To do this,  we first assumed similarly as in previous studies \citep{condon_14_1989,chiaberge_low-power_2009,castignani_cluster_2014} a power-law for the radio spectrum, that is,  $S_\nu\propto\nu^{-\alpha}$,  where $S_\nu$ is the radio flux density at the observer frequency $\nu,$ and the spectral index $\alpha$ is fixed to $\alpha=0.8$. We then converted the 1.4~GHz VLA radio fluxes $S_{\rm 1.4~GHz}$ into rest-frame 1.4 GHz luminosity densities as follows:

\begin{equation}
\label{eq:L14formula} 
L_{1.4~{\rm GHz}} = 4\pi\;S_{1.4~{\rm GHz}}\;D_L(z)^2\left(1+z\right)^{\alpha-1},
\end{equation}
where D$_L$ is the luminosity distance.

Figure~\ref{fig:WISEcolors_radio_luminosity} (top) displays our sources in the $L_{\rm 1.4~GHz}$ versus redshift plane.
They all have $L_{1.4~{\rm GHz}}\gtrsim3\times10^{30}~~{\rm erg}~{\rm s}^{-1}~{\rm Hz}^{-1}$ , which is typical of radio-loud AGN, while purely starburst galaxies have lower $L_{1.4~{\rm GHz}}<10^{30}~~{\rm erg}~{\rm s}^{-1}~{\rm Hz}^{-1}$ \citep{chiaberge_low-power_2009}.

Furthermore, the majority (71\%, i.e., 30/42) of our sources have high radio powers, greater than $L_{\rm 1.4~GHz}=2\times10^{32}~{\rm erg}~{\rm s}^{-1}~{\rm Hz}^{-1}$, which we used to distinguish between low-luminosity radio sources (LLRS) and high-luminosity radio sources (HLRS), similarly as in previous studies \citep[e.g.,][]{chiaberge_low-power_2009,castignani_cluster_2014}.
As the radio galaxy population has a bimodal distribution in radio power, it is worth mentioning that the adopted LLRS to HLRS luminosity threshold corresponds to the fiducial radio power that fairly separates Faranoff-Riley~I (FR~I) from FR~II radio galaxies \citep[][]{fanaroff_morphology_1974,zirbel_properties_1996}.
Furthermore, as a result of the Malmquist bias associated with the VLA FIRST flux limit of $\sim1$~mJy, the fraction of HLRSs increases with redshift and reaches unity at $z>1$.

Figure~\ref{fig:WISEcolors_radio_luminosity} (bottom) shows the sources in our sample in the WISE color-color diagram, where sources are distinguished using the color-based classification by \citet{jarrett_galaxy_2017}, as highlighted in the figure. Interestingly, our sample populates only three regions in the diagram. The majority (28 out of 42) of our sources are classified as AGN based on WISE colors. This is expected because they were selected as distant and powerful radio sources at $z>0.3$. Based on WISE colors, the remaining sources are fairly equally distributed between the intermediate disk (9 out of 42) and starburst  (5 out of 42) classes.

Furthermore, as shown in Fig.~\ref{fig:WISEcolors_radio_luminosity} (top), the WISE IR colors of the vast majority of $z>1$ sources are consistent with AGN contribution. They also show high 1.4~GHz radio luminosities typical of radio-loud quasars (QSOs). The majority (22 out of 42, i.e., 52\%) of our sources are in fact classified as quasars in the NASA/IPAC extragalactic database (NED), for instance, with counterparts in the 2dF–SDSS LRG And QSO \citep[2SLAQ;][]{croom_2dfsdss_2009} catalog, or with X-ray conterparts \citep[XMM;][]{rosen_xmm-newton_2016}, as outlined in Table~\ref{tab:A1sample}.

\section{Black hole, jet, accretion, and stellar properties}\label{sec:BH_jet_Mstar}
\subsection{Black hole masses}
One of the main goals of this work is to investigate the coevolution of central black holes with the host galaxies of the radio-loud AGN in our sample. To do this, we estimated black hole masses using the widely used  single-epoch (SE) method, which is particularly suited for distant type~1 AGN. According to this method, black hole masses $M_{\rm BH}$ can be estimated under the assumption that the broad-line region (BLR) is in virial equilibrium, as follows:

\begin{equation}
M_{\rm BH} = f\;\frac{R_{\rm BLR}\Delta V^2}{G}\;,
\end{equation}

where $R_{\rm BLR}$ is the BLR radius, $\Delta V$ is the velocity of the BLR clouds that can be estimated from the broad emission line width, $f$ is the virial coefficient that depends on the geometry and kinematics of the BLR, and G is the gravitational constant. The SE method then uses the relation between the BLR size and
the AGN optical/UV continuum luminosity empirically found from reverberation mapping \citep{peterson_central_2004,kaspi_reverberation_2007,bentz_lick_2009}, as well as the tight correlation between the continuum luminosity and that of broad emission lines \citep[e.g.,][]{shen_catalog_2011}. 
With these considerations, the black hole mass can be expressed as

\begin{equation}
\label{eq:MBH}
\log \left( \frac{M_{\rm BH}}{M_{\odot}} \right) = a + b\;\log \left( \frac{L}{10^{44}~{\rm erg~s}^{-1}} \right) + c\;\log \left( \frac{\rm FWHM}{{\rm km~s}^{-1}} \right)\;,
\end{equation} 

\noindent where the coefficients $a$, $b$, and $c$ are empirically calibrated against local AGNs with reverberation mapping masses or using different lines. $L$ and FWHM are the line luminosity and width.  We used the coefficients obtained for the \halpha, \hbeta, \mgii, and \civ~ broad emission lines by \citet{shen_catalog_2011} and \citet{shaw_spectroscopy_2012}. These are widely used lines that are redshifted in the optical domain, depending on the redshift of the source. These lines indeed enable estimates of black hole masses over a wide range of redhifts. Similarly to previous studies \citep{shaw_spectroscopy_2012,castignani_black-hole_2013}, we used \halpha, \hbeta, and \mgii~ for sources at $z < 1$ and the \mgii~ and \civ~ lines for sources at higher redshifts. 
When multiple broad emission lines were available for a given sources, we adopted the following order of preference: \halpha, \hbeta, \mgii, and \civ~(see, e.g., \cite{shen_comparing_2012} for more details).
In Table~\ref{tab:coeff} we report the coefficients used in this work, and in Table~\ref{tab:A2MBH} we list the black hole masses.

\begin{figure*}
\begin{center}
\captionsetup[subfigure]{labelformat=empty}
\subfloat[]{\includegraphics[trim=0cm 0cm 0cm 0cm, clip=true, width=0.4\textwidth]{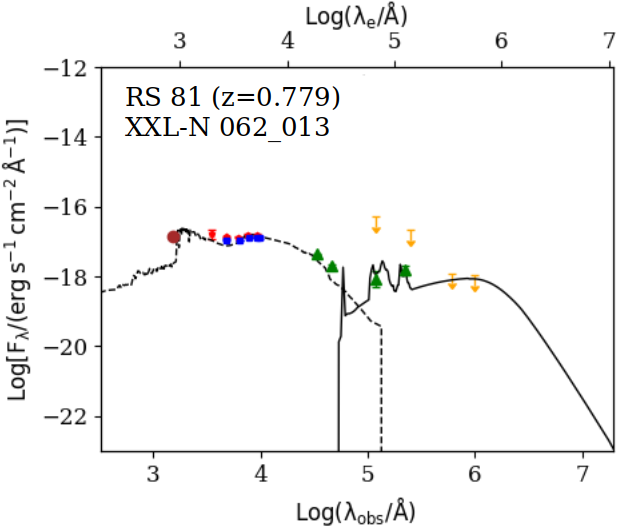}}
\hspace{0.5cm}\subfloat[]{\includegraphics[trim=0cm 0cm 0cm 0cm, clip=true, width=0.4\textwidth]{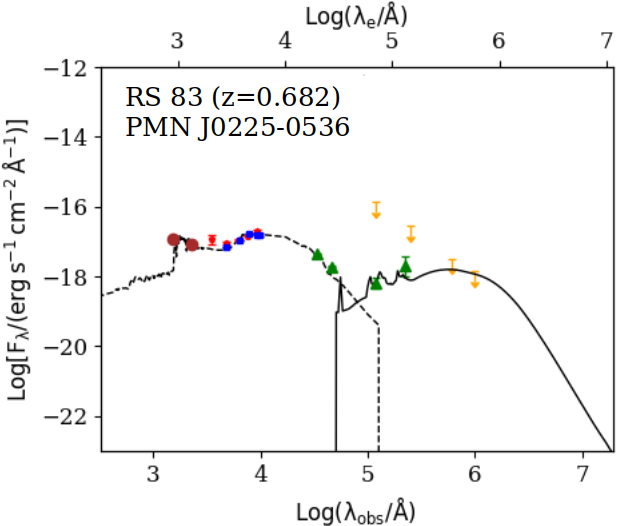}} \\
\vspace{-0.5cm}
\subfloat[]{\includegraphics[trim=0cm 0cm 0cm 0cm, clip=true, width=0.4\textwidth]{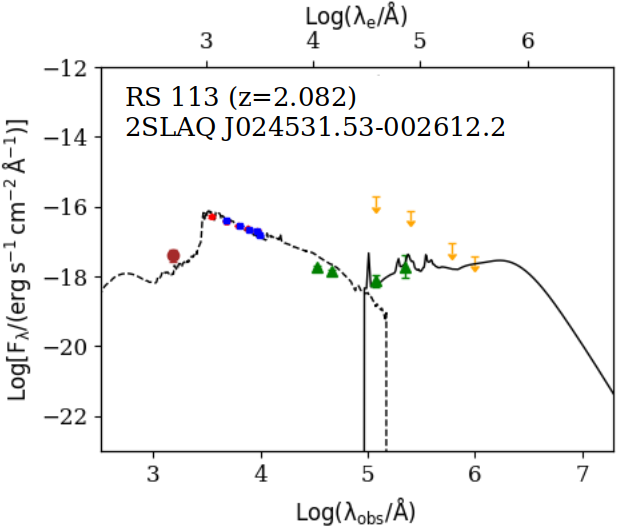}}
\hspace{0.5cm}\subfloat[]{\includegraphics[trim=0cm 0cm 0cm 0cm, clip=true, width=0.4\textwidth]{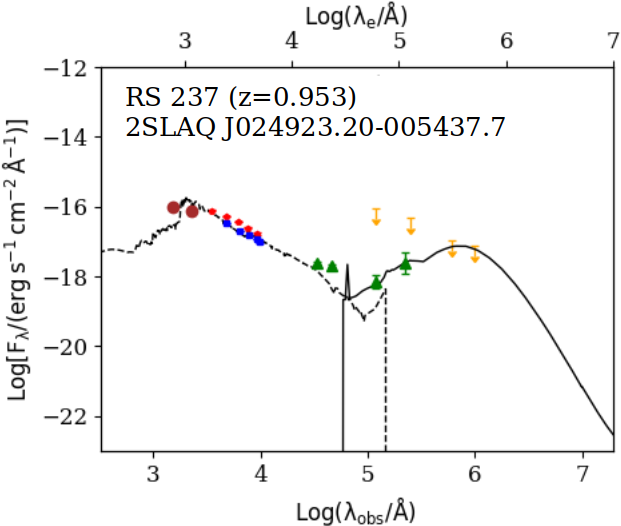}}
\caption{Examples of SEDs of the radio  sources in our sample with black hole mass estimates. The IDs, names, and redshifts of the galaxies are shown at the top left of each panel. Data-points are from GALEX (brown dots), SDSS (red pentagons), DES (blue squares), WISE (green triangles), and IRAS (yellow upper limits). Dashed and solid lines are the best-fit models for the stellar and dust components, respectively.}
\label{fig:SEDs}
\end{center}
\end{figure*}

\begin{table}
        \begin{center}
    \begin{tabular}{|c|ccc|}
                \hline 
                $ $ & $a$ & $b$ & $c$ \\
                \hline
                %%%%%%%%%%%%%%%%%%%%%%%%
                \halpha~ & $1.24$ & $0.43$ & $2.1$\\
                \hbeta~ & $1.63$ & $0.49$ & $2.0$ \\
                \mgii~ & $1.70$ & $0.63$ & $2.0$ \\
                \civ~ & $1.52$ & $0.46$ & $2.0$ \\
                \hline
        \end{tabular}
        \end{center}
        \caption{Coefficients for estimating black hole masses using broad emission lines and Eq.~\ref{eq:MBH}. Values for the \hbeta, MgII, and CIV lines are from \citet{shaw_spectroscopy_2012}. 
        The coefficients for \halpha~ come from \citet{shen_catalog_2011}}
        \label{tab:coeff}
\end{table}

\subsection{Jet power}\label{sec:jet_power}
The sources in our sample are radio-loud AGN that are typically characterized by jetted outflows that strongly emit at radio wavelengths mainly via synchrotron emission. By studying jet properties such as its total power, we investigate the complex interplay between the jet, the black hole, and the gas accretion onto it, which is commonly referred to as radio-mode AGN feedback \citep[e.g.,][for a review]{fabian_observational_2012}.
Following previous work by \citet{willott_emission_1999}, we estimate the jet power as
\begin{equation}
    \label{eq:Qjet}
        Q_{jet} = 3\times10^{45}\;\xi^{3/2} \left(\frac{L_{\rm 151~MHz}}{10^{35}~{\rm erg~s^{-1}~Hz} ^{-1}~{\rm sr}^{-1}}\right)^{6/7} \ {\rm erg~s^{-1}}\;,
\end{equation}
\noindent where $L_{\rm 151~MHz}$ is the extended total radio luminosity density at 151~MHz in the rest frame, and $\xi$  ranges between 10 and 20. We used an intermediate value $\xi=15$. To estimate $L_{\rm 151~MHz}$ , we extrapolated the $L_{\rm 1.4~GHz}$ luminosity densities assuming an isotropic emission and a power law with $\alpha=0.8$, as further described in Sect.~\ref{sec:radio_IR_prop}. The resulting jet powers are reported in Table~\ref{tab:A2MBH}.

\subsection{Eddington ratio}\label{sec:Edd_ratio}
We wish to link the gas accretion onto the black hole with the AGN properties. We therefore estimated the Eddington ratio, which is defined as\
\begin{equation}
\eta = \frac{L_{\rm disk}}{L_{\rm Edd}}\;,
\end{equation}
where $L_{\rm disk}$ and $L_{\rm Edd}$ are the disk and Eddington luminosities. The latter can be expressed as
\begin{equation}
L_{\rm Edd}=1.26\times10^{38} \left(\frac{M_{\rm BH}}{M_{\odot}}\right)~{\rm erg~s}^{-1}\;.
\end{equation}

To estimate the disk luminosity, we instead followed the prescriptions described in \cite{celotti_jets_1997}.
First, we assumed that BLR contributes $\sim10\%$ of the total disk luminosity, that is, $L_{\rm disk} \simeq 10\;L_{\rm BLR}$. To estimate the BLR luminosity, we then used the line ratios reported in \cite{francis_high_1991}, which are typical line luminosities of bright QSOs, normalized to that of the Ly$\alpha$ emission. The BLR luminosity can therefore be estimated as 

\begin{equation}
L_{\rm BLR}=<L_{\rm BLR}^*> \frac{\sum_{i}^{} L_{i,{\rm obs}}}{\sum_{i}^{} L_{i,{\rm est}}^*}\;,
\end{equation}

\noindent where $L_{i,{\rm obs}}$ is the luminosity of the observed $i$th line in the BLR, and $L_{i,{\rm est}}^*$ is the line ratio of the $i$th line presented in the table of \citet{francis_high_1991}. With these prescriptions, the total normalized BLR luminosity is equal to  $<L_{BLR}^*>= L_{\rm H\alpha}^* + L_{\rm H{\beta}}^* + L_{{\rm C}{~\mathsc{iv}}}^* + L_{{\rm Mg}~{\mathsc{ii}}}^* +  L_{\rm Ly\alpha}^* +  L_{\rm Ly\beta}^*  + L_{\rm H\gamma}^* + L_{\rm Al\;\mathsc{III}}^* + L_{\rm Si~\mathsc{IV}}^* + L_{\rm C~\mathsc{II}}^*  + L_{\rm O~\mathsc{I}}^* = 390.3$, where $L_{\rm Ly\alpha}^*=100$ was fixed as a reference.

The resulting Eddington ratios are reported in  Table~\ref{tab:A2MBH}. They are mostly in the range $\log\eta\sim(-4;-1)$, with a median = -1.9, as typically found for type~1 radio-loud AGN, but lower than the ratios of type~2 quasars  \citep[e.g.,][]{castignani_black-hole_2013,kong_black_2018}.

\subsection{SED modeling}
The radio sources in our sample have a broad multiwavelength photometric coverage from the UV to the IR, which enables the determination of stellar masses and SFR estimates via SED modeling.

For the GAMA sources in our sample, we considered the SED fits by \citet{driver_gamag10-cosmos3d-hst_2018} performed with MAGPHYS \citep{da_cunha_simple_2008}. Photometric data include GALEX \citep{martin_galaxy_2005,morrissey_calibration_2007} in the UV, SDSS \citep{york_sloan_2000} in the optical, as
well as the VISTA Kilo-degree IR Galaxy Survey 
\citep[VIKING,][]{edge_vista_2013}, WISE \citep{wright_wide-field_2010}, and {\it Herschel}-
ATLAS \citep{eales_herschel_2010,valiante_herschel-atlas_2016} in the near- to far-IR.

For the sources in the DES SN deep fields, available photometry includes GALEX in the UV, \textsf{ugriz} (SDSS) and \textsf{grizY} (DES) magnitudes in the optical, WISE data in the near-IR, as well as IRAS upper limits in the far-IR, which we gathered as in \citet{castignani_molecular_2019}, to which we refer for further details. In this previous work, we followed up two radio sources in molecular gas in dense megaparsec-scale environments at $z=0.4$ and $z=0.6$ within the DES SN deep fields, while we enlarge the sample here to investigate the coevolution of black holes with radio sources.

Analogously to \citet{castignani_molecular_2019}, we then performed fits to the SEDs using LePhare \citep{arnouts_measuring_1999,ilbert_accurate_2006}. Following the prescriptions provided for the LePhare code, we fit the far-IR data separately to account for
possible dust emission, using the \citet{chary_interpreting_2001} library
consisting of 105 templates. The remaining photometric data points at shorter wavelengths were fit using the CE\_NEW\_MOD library, which consists of 66 galaxy templates based on linear interpolation of the four original SEDs of \citet{coleman_colors_1980}. We then converted the rest frame (8.0-1000)~$\mu$m IR (dust) luminosity into an SFR estimate by using the \citet{Kennicutt1998} relation, calibrated to a \citet{Chabrier2003} initial mass function. The SEDs of four of our radio sources are shown in Fig.~\ref{fig:SEDs}. RS~81 and 83 have prominent elliptical emission in the optical domain, while their IR emission is consistent with dust emission due to star formation. They are indeed classified as intermediate disks based on WISE colors. 
RS~113 and 237 are WISE AGN and show steep SEDs at near-IR to optical wavelengths, which suggests that the emission is contaminated by nonthermal AGN emission. 
\begin{figure}
    \centering
    \hspace*{-0cm}
    \includegraphics[width=1\linewidth]{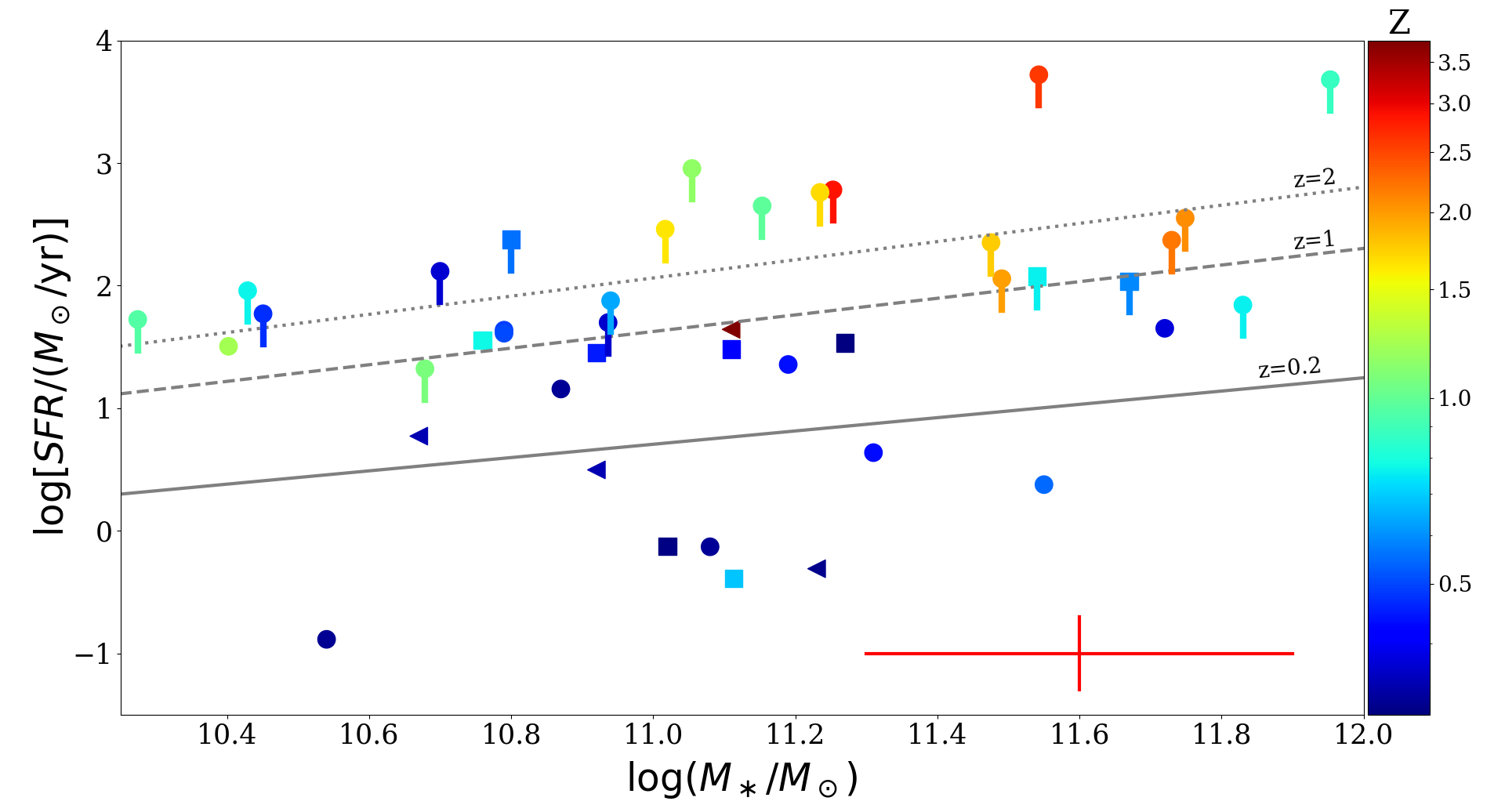}               
    \caption{Plot of the SFR vs. stellar mass for the galaxies in our sample. Sources are color-coded according to their redshift, while the different symbols correspond to the different WISE classes (AGN as circles, starbursts as triangles, and intermediate disks as squares), as in Fig.~\ref{fig:WISEcolors_radio_luminosity}. Upper limits to the SFRs are indicated with arrows. The diagonal lines correspond to the MS model prescriptions by \citet{Speagle2014} at $z=0.2$, 1, and 2. The red cross in the bottom right corner shows the typical uncertainties of $\sim$0.3~dex for both SFRs and stellar masses.}
    \label{fig:SFR-M*}
\end{figure}

\subsection{SFR versus stellar mass}\label{sec:SFR_vs_Mstar}
Figure~\ref{fig:SFR-M*} displays the radio sources of our sample in the SFR versus stellar mass ($M_\star$) plane resulting from the SED fits. The sources are color-coded according to the redshift, while the different symbols correspond to the WISE classification.

Overall, the sources are massive, with $\log(M_\star/M_\odot)\simeq10.3-12.0$ (median=10.1), which agrees with them being radio-loud AGN, which are indeed typically hosted by massive ellipticals
\citep{best_host_2005}. Our galaxies also mostly lie along the star-forming main sequence  (MS), although with a large scatter. The mean specific SFR is sSFR=0.40$\pm$0.44~Gyr$^{-1}$, where the root mean square (rms) dispersion is reported as uncertainty.

Galaxies at higher redshifts tend to have higher SFRs, in agreement with the MS model prescriptions \citep{speagle_highly_2014}. However, as highlighted in Sect.~\ref{sec:radio_IR_prop}, the fraction of sources with AGN contamination also increases with redshift, which may result in biased-high SFRs. The latter may be the case where the optical-IR SED is steep, and thus the IR emission is likely dominated by the AGN contribution, more than star formation.
To overcome this limitation, we conservatively reconsidered the SFR estimates and assigned upper limits when the SFRs largely exceeded 100~$M_\star$/yr or in the cases of steep-spectrum SEDs (e.g., for RS~113 and 237 mentioned above). Namely, we considered as steep spectra those AGN whose optical-IR SED has a characteristic power-law behavior  $F_\lambda\propto\lambda^{-1}$. We verified \textit{\textup{a posteriori}} that these radio sources are indeed mostly located in the upper part of the MS and are classified as WISE AGN.

\section{Comparison sample}
\label{sec:control}

To place the AGN in our sample into context, we additionally considered a compilation of sources with available black hole and stellar mass estimates.  
We first took the 30 nearby galaxies from \citet{haring_black_2004}. Galaxy masses were derived by the authors through Jeans equation modeling or were adopted from dynamical models in the literature, and black hole masses are from \citet{tremaine_slope_2002} and references therein.
Then, we added the 35 nearby galaxies from the sample of \cite{mcconnell_revisiting_2013}, who expanded and revised  available galaxy bulge masses and dynamical measurements of black hole masses.
\citet{cisternas_secular_2011} has 32 type~1 AGN at $z=0.3-0.9$, drawn from the XMM-COSMOS survey. Available stellar masses are based on the modeling of {\it HST} images, taking both AGN and host galaxy contributions into account, and  black hole masses are from \hbeta~ \citep{trump_nature_2009}.
We also added the 18 broad-line X-ray AGN $0.5<z<1.2$ in the Extended {\it Chandra} Deep Field-South Survey from \cite{schramm_black_2013}, who estimated \mgii -based black hole masses, as well as {\it HST} color-based stellar mass estimates.
We included 78 radio-quite type 1 AGN at $z\simeq1-2$ in the COSMOS survey \cite{merloni_cosmic_2009}. Stellar masses were determined via SED fitting, and black hole masses are based on the \mgii~ emission lines of VIMOS/VLT spectra. 
The 10 type~1 AGN at $1<z<2$ reported in COSMOS from \citet{jahnke_massive_2009}, who estimated {\it HST} color- based stellar masses, and the virial black hole masses come from the spectroscopic COSMOS Magellan/IMACS and zCOSMOS surveys.
We added the 53 radio-quiet QSOs at $z<3$ from \citet{decarli_quasar_2010,decarli_quasar_2010-1}. Virial black hole masses come from the \hbeta, \mgii, and \civ~ emission lines, and the stellar masses were estimated by the authors assuming a stellar R-band mass-to-light ratio.
\citet{targett_host_2012} gives us the information about 2 SDSS luminous quasars at $z\sim4$. Virial black hole mass estimates come from \civ~ emission, and the stellar masses were estimated on the basis of the evolutionary synthesis models of \citet{bruzual_stellar_2003}.
\cite{shields_black_2006} reported 9 distant $z\sim1.4-6.4$ QSOs. Black hole masses were derived from broad emission lines, and they used CO emission line widths to infer the dynamical bulge masses. 
Finally, we added the 7 QSOs  at $z\simeq 6$ from \cite{wang_molecular_2010}. They calculated the stellar mass as the difference between the bulge dynamical mass and the CO molecular gas mass. For these QSOs, we used the black hole masses  adopted by the authors that were estimated using the AGN continuum luminosity \citep{Jiang2006,Wang2008}.

In addition to the sources listed above, the second group of galaxies that we used as a comparison is composed by powerful AGN with available estimates of the black hole mass, jet power, and   Eddington ratio.
We added the 44 radio-loud AGN studied in \citet{le_jet_2018} at redshifts $z<0.2$, {which is lower than those of the radio sources in our sample.} These sources have estimates of the jet powers \citep{le_jet_2018} and of their black hole masses \citep{Allen2006,balmaverde_accretion_2008}.
We also added the 208 $\gamma$-ray {\it Fermi} blazars at $0<z<3.1$  from \cite{xiong_intrinsic_2014}. Virial black hole mass estimates mostly come from different broad emission lines, and the rest were obtained from scaling relations. Jet powers $Q_{\rm jet}$ are mostly from \citet{nemmen_universal_2012} and were estimated using the correlation between the extended radio emission and the jet power. Alternatively,  \citet{xiong_intrinsic_2014} calculated $Q_{\rm jet}$ using the scaling relation provided by \citet{nemmen_universal_2012} between the $\gamma$-ray luminosity and the kinetic power.
Finally, \citet{liu_jet_2006} has 146 radio-loud QSOs at {$0.1<z<2.5$},  classified as flat-spectrum (54\%) or steep-spectrum radio quasars (46\%).  The black hole virial masses come from the \hbeta, \mgii, or \civ~ emission lines. The jet power were calculated by the authors using low-frequency radio emission, following \citet{punsly_independent_2005}.

{These sources outlined above are radio-loud AGN. However, we verified that none of them are included in our sample. While these studies investigated the black hole and jet properties of large samples of radio sources, they did not characterize their IR to optical SEDs, as we did here for our smaller sample of radio sources.}

\begin{figure}[h!]
        \centering
                %\hspace{-1.0cm}
        \includegraphics[trim={14cm 0 0 0},clip,scale=0.36]{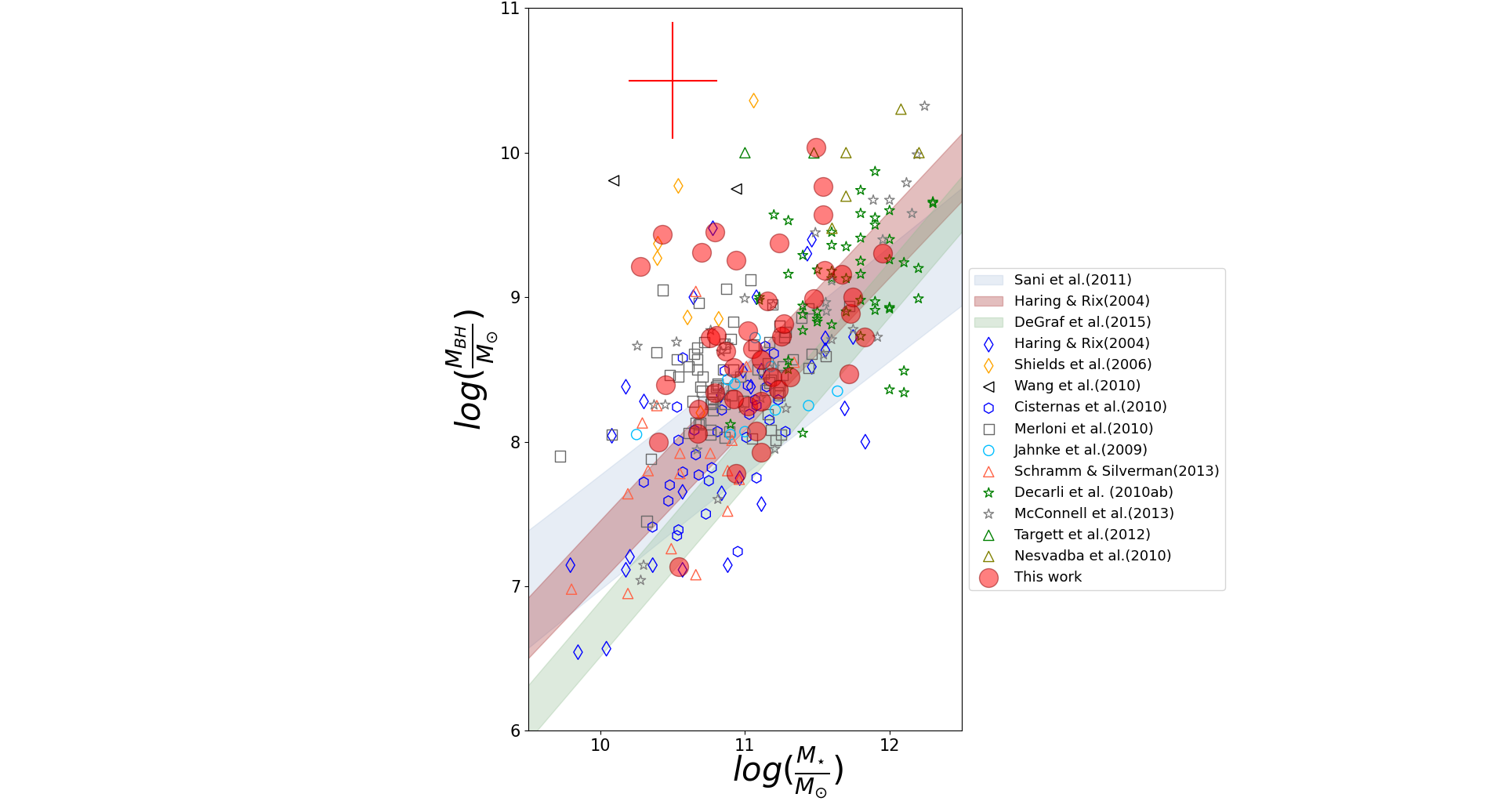}
        \caption{Scatter plot of black hole vs. stellar mass. Filled red dots correspond to our sample of radio sources, and the sources in the comparison sample are shown as open symbols. Scaling relations are overlaid \citep{sani_spitzerirac_2011,degraf_scaling_2015,haring_black_2004}. The legend at the right displays the adopted color code. The red cross in the top left corner shows the typical uncertainties $\sim$0.4~dex and $\sim$0.3~dex for the black hole and stellar masses, respectively.}
        \label{fig:MBH-M*}
\end{figure}

\begin{figure*}[t!]
        \hspace{-0cm}
                \centering
                \includegraphics[width=1.0\linewidth]{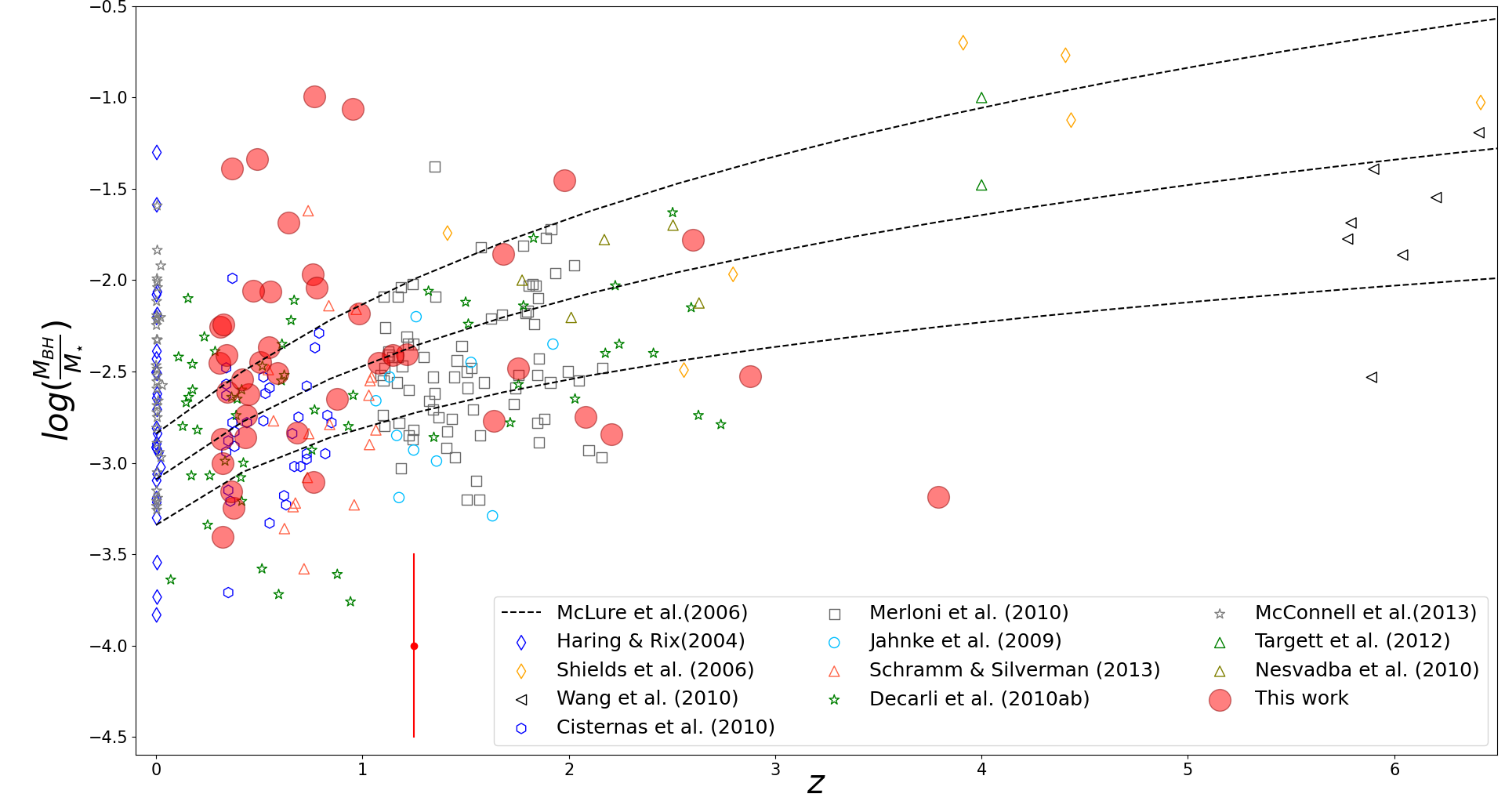}
                \caption{Black hole to stellar mass ratio as a function of redshift for our sample of radio sources (filled red dots) and for the galaxies in our comparison sample (open symbols). The dashed black lines correspond to the evolutionary model described by \citet{mclure_evolution_2006}, along with the associated $1\sigma$ uncertainty. The color-coding is reported in the legend in the bottom right corner. The error bars to the left of the legend show the typical $\sim0.5$~dex uncertainty in $\log(M_{\rm BH}/M_\star)$.}
                \label{fig:MBH-M*-z}
\end{figure*}

\section{Results}
\label{sec:results}

In this section, we report different scaling relations for the radio sources in our sample, including black hole and stellar masses, jet powers, Eddington ratios, and the redshifts.  We also include the sources from the literature as outlined in Sect.~\ref{sec:control} as a comparison, as well as the scaling relations derived in previous studies.

\subsection{Black hole versus stellar mass relation}\label{sec:BH_vs_Mstar}
We start by considering black hole and stellar masses and their relative evolution with redshift. Figure~\ref{fig:MBH-M*} displays the black hole mass ($M_{\rm BH}$) versus the stellar mass ($M_\star$).  Interestingly, our radio-loud sources nicely follow the trend previously observed for different samples of both local galaxies and distant AGN, overplotted in the figure, and for those inferred by the scaling relations, which were also reported \citep{sani_spitzerirac_2011,haring_black_2004,degraf_scaling_2015}.

In particular, our sources populate the high $\log(M_{\rm BH}/M_\odot)\simeq7.1-10.0$ and high $\log(M_\star/M_\odot)\simeq10.2-12.0$ region in Fig.~\ref{fig:MBH-M*} densely, which agrees with the fact that radio-loud AGN are almost invariably associated with the most massive galaxies and black holes.  Interestingly, a substantial fraction of our sources 9/42 (i.e., 21\%) have black hole masses $\log(M_{\rm BH}/M_\odot)>9$ well above the scaling relations for both local \citep{haring_black_2004,sani_spitzerirac_2011} and distant sources at the median redhift $z=0.6$ of our sample \citep{degraf_scaling_2015}. This behavior suggests that the growth of black hole masses in radio-loud AGN largely occurs at early $z>1$ epochs, while the early stellar mass assembly may not be equally effective. Substantial growth of the stellar mass may take place even at lower redshifts in order to flatten the observed 
$M_{\rm BH}$-$M_\star$ scaling relation by $z=0$. Previous studies indeed suggested that massive ellipticals may double their stellar mass between $z=1$ and $z=0$ \citep{Ilbert2010,Lidman2012}.

We further investigate this evolutionary scenario in Fig.~\ref{fig:MBH-M*-z}, which shows the $M_{\rm BH}/ M_\star$ ratio as a function of redshift.
The large majority (36 out of 42, i.e., 86\%) of our radio sources have $M_{\rm BH}/M_\star$ ratios that are similar to those of AGN in the comparison sample at similar redshifts,  and they agree with model prescriptions by \citet{mclure_evolution_2006}, which are overplotted as dashed lines in Fig.~\ref{fig:MBH-M*-z}.

 It is worth mentioning that our sample of radio sources is flux limited. However, we expect the Malmquist bias to have a marginal impact on the $(M_{\rm BH}/ M_\star)$ ratio.  $M_{\rm BH}$ and $M_\star$ both scale with the BLR line and the IR to optical luminosities, respectively, and therefore have a similar dependence on redshift via the luminosity distance. Furthermore, as shown in Fig.~\ref{fig:MBH-M*-z}, at fixed redshift, the $M_{\rm BH}/ M_\star$ ratios of both our radio sources and those in the comparison sample span a broad range. Similarly, when plotting $M_\star$ and $M_{\rm BH}$ against redshift separately, we did not find any clear trend, as indeed the points are scattered at fixed redshift. These findings suggest that any possible Malmquist bias likely has a subdominant effect.

There are, however, five clear outliers among our radio sources. RS~197 at the highest redshift $z=3.79$  has a low $\log(M_{\rm BH}/M_\star)=-3.2$, which is well below the expected range of values, according to the \citet{mclure_evolution_2006} model prescription. Furthermore, a substantial fraction (five sources, i.e., 12\%) of our radio loud AGN, namely RS~214, RS~237, G~537618, G~721940, and G~746605, at redshifts between $z\sim0.37-0.95$, have high $\log M_{\rm BH}/ M_\star$ ratios in the range $\sim(-1.69;-1.0)$, which is well above the model predictions displayed in Fig.~\ref{fig:MBH-M*-z}, as well as higher than the ratios found in AGN at similar redshifts. Indeed, in the redshift range $z\sim0.3-3.8$ spanned by our radio sources, there are only 3 out of 186 (i.e., 1.6\%) AGN with $\log M_{\rm BH}/M_\star > -1.69$ in our comparison sample, while the proportion is significantly higher (12\%) for our radio-loud AGN.
These results suggest that a non-negligible fraction of radio-loud AGN  may experience a different stellar mass assembly path than radio-quiet AGN. We stress that these five radio sources are a subsample of the nine outliers of Fig.~\ref{fig:MBH-M*}, as discussed above, and have high S/N  line fluxes in \hbeta~ or \mgii, which yielded robust $M_{\rm BH}$ estimates. The only exception is represented by G~721940, for which the \hbeta~ emission and associated FWHM are at lower S/N$\sim2$, as highlighted in Table~\ref{tab:A2MBH}.

The excess of overmassive SMBHs in radio-loud AGN suggests that their stellar and SMBH mass buildup is regulated by their large-scale radio jets. A possible scenario is that SMBHs of the subpopulation with high $M_{\rm BH}/M_\star$ are mature, that is, their mass has been effectively assembled already by redshift $z=1$ via accretion \citep[e.g.,][]{delvecchio2018}. On the other hand, their stellar mass growth may not have occurred as effectively as in the overall AGN population, plausibly  because of radio-mode AGN feedback \citep{fabian_observational_2012}. While the accretion of hot gas onto the SMBH sustains the AGN activity and the SMBH growth, the large scale radio jets may prevent accretion and cooling of the inter-galactic medium gas, which is ultimately responsible for the stellar mass assembly. Altogether, we suggest that radio-mode AGN feedback results in the observed high values for $M_{\rm BH}/M_\star$ in radio-loud AGN.

In order to investigate further this scenario, we link accretion and jet properties to the black hole mass in the next sections by considering both the jet power and Eddington ratio of our radio sources. We stress that the usual $M_{\rm BH}-M_{\rm bulge}$ or $M_{\rm BH}-\sigma$ relations typically refer to the central spheroid and not to the total stellar mass
\citep[e.g.,][]{kormendy_coevolution_2013}.  However, our radio-loud AGN sample is composed in a large majority of early-type galaxies, where the spheroid constitutes most of the stellar mass, and this approximation is justified. 
Furthermore, because of the potential AGN contamination to the SED, the stellar mass may be biased high. This implies that $M_{\rm BH}/M_\star$ ratios can be even higher than reported. By considering $M_{\rm BH}/M_\star$ ratios as lower limits, we would have an even  stronger discrepancy, in particular, for the subsample of high $M_{\rm BH}/M_\star$ radio sources mentioned above, with respect to the model prescriptions and the comparison sample of distant AGN at fixed resdshift. All these results seem to corroborate the scenario that SMBH growth is more rapid than stellar mass assembly, and this is particularly true for distant radio sources, in comparison to the overall AGN population.

\subsection{Jet power, black hole mass, and accretion}
\begin{figure}
\begin{center}
        \hspace{-0cm}
        \includegraphics[scale=0.25]{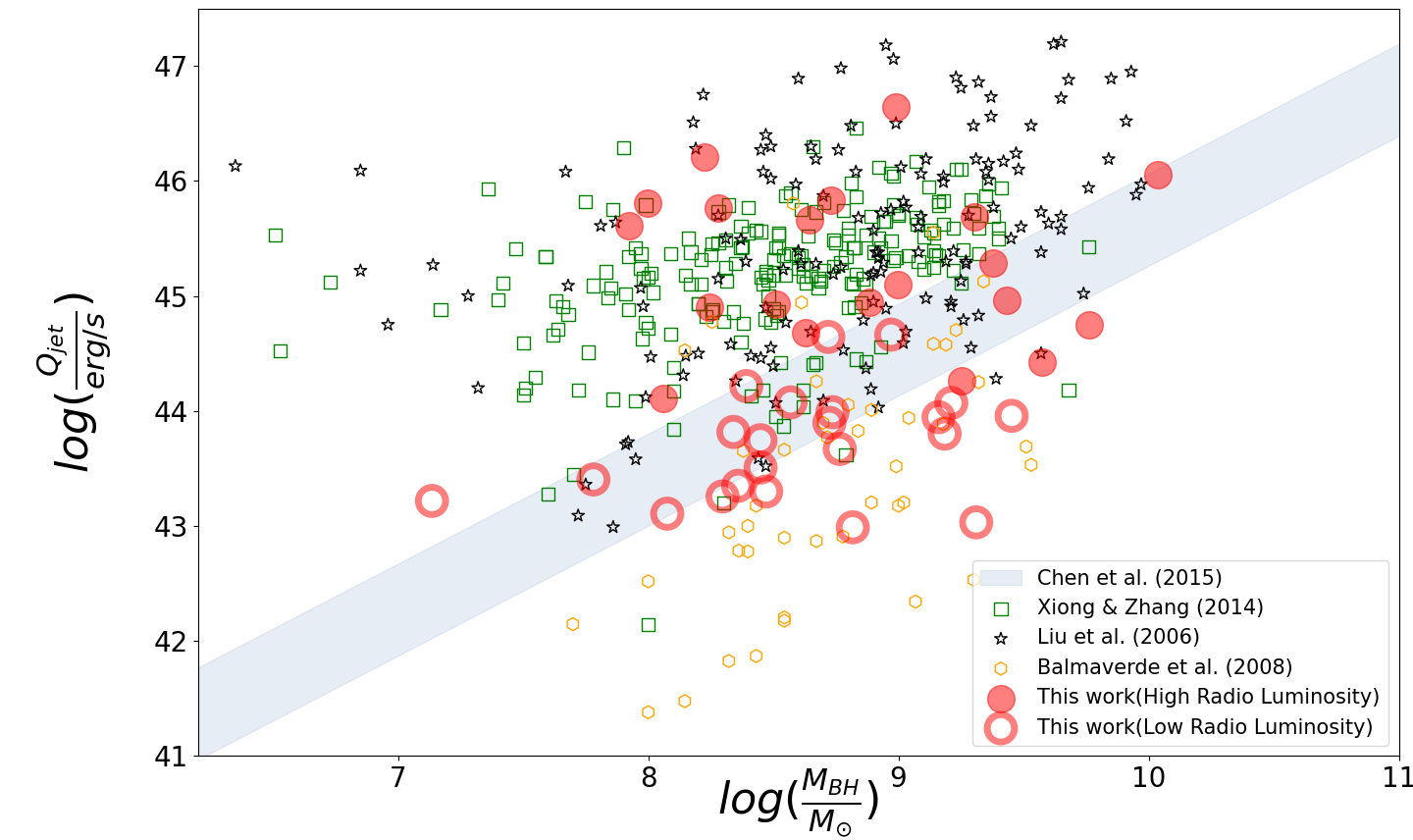}
        \caption{Jet power vs. black hole mass. The diagonal dashed region represents the model reported in previous studies \citep{Wu_Cao2008,Chen2015} that distinguishes between FR~I and FR~II radio galaxies. Sources from our sample are distinguished between high- and low-luminosity radio sources. Sources from our comparison sample are also shown. We refer to the legend for the color-code.}
        \label{fig:Qjet-MBH}
        \end{center}
\end{figure}

As mentioned in the previous sections, mechanical radio-mode AGN feedback can regulate the cooling of hot gas in the intergalactic medium, and thus the stellar mass growth of the host galaxy itself as well as the accretion onto the central SMBH. To better understand the interplay between jet, black hole, and accretion properties, in Fig.~\ref{fig:Qjet-MBH} we show the jet power $Q_{\rm jet}$  (see Sect.~\ref{sec:jet_power}), plotted against the black hole mass $M_{\rm BH}$. The radio sources of our sample are highlighted, and we also overplot the comparison sources 
outlined in Sect.~\ref{sec:control} \citep{liu_jet_2006,balmaverde_accretion_2008,xiong_intrinsic_2014,Chen2015}. 

Our radio-loud AGN densely populate the upper right region of the  $Q_{\rm jet}$-$M_{\rm BH}$ plane, which is occupied by sources with high values of both the black hole mass ($M_{\rm BH}\gtrsim10^8M_\odot$) and the jet power ($Q_{\rm jet}\gtrsim10^{43}$~erg~s$^{-1}$). Sources in the comparison sample similarly occupy this region, while they also extend to lower values of $M_{\rm BH}$ \citep{liu_jet_2006,xiong_intrinsic_2014} and jet power \citep{balmaverde_accretion_2008}. 
{These results suggest that the distant radio-loud AGN, quite independently of the redshift, are almost invariably associated with massive black holes and powerful radio jets. This agrees with the tight connection existing between black hole accretion and jet production in powerful radio-loud AGN \citep[e.g.,][]{Ghisellini2014,Sbarrato2014,Inoue2017}.}

Furthermore, HLRSs are characterized by a jet power that is typically higher than in LLRSs.
As discussed in Sect.~\ref{sec:radio_IR_prop}, these two classes indeed have 1.4~GHz rest-frame luminosity densities typical of FR~II and FR~I radio galaxies, respectively. As $Q_{\rm jet}$ increases with the radio luminosity density (Eq.~\ref{eq:Qjet}), high-luminosity radio sources have higher $Q_{\rm jet}$ values than low-luminosity sources. Furthermore, the two classes of LLRGs and HLRGs are also delimited in the $Q_{\rm jet}$-$M_{\rm BH}$ plane by the relation found in previous studies \citep{Wu_Cao2008,Chen2015}, originally used to distinguish between FR~I and FR~II radio galaxy populations. The clear separation of LLRGs and HLRGs in the $Q_{\rm jet}$ versus $M_{\rm BH}$ plane can be explained by combining the $M_{\rm BH}$ versus  $M_{\rm bulge}$ relation for elliptical galaxies and the relation between $Q_{\rm jet}$ and the host galaxy optical luminosity  \citep{Ledlow_Owen1996} that separates the FR~I and FR~II radio galaxies. The combination of these two relations also implies the observed spread of our sources in Fig.~\ref{fig:Qjet-MBH}. We did not find any significant correlation (as measured with the Spearman test) between $Q_{\rm jet}$ and $M_{\rm BH}$ for our radio sources.

\begin{figure}
\begin{center}
        \hspace{0cm}
        \includegraphics[scale=0.25]{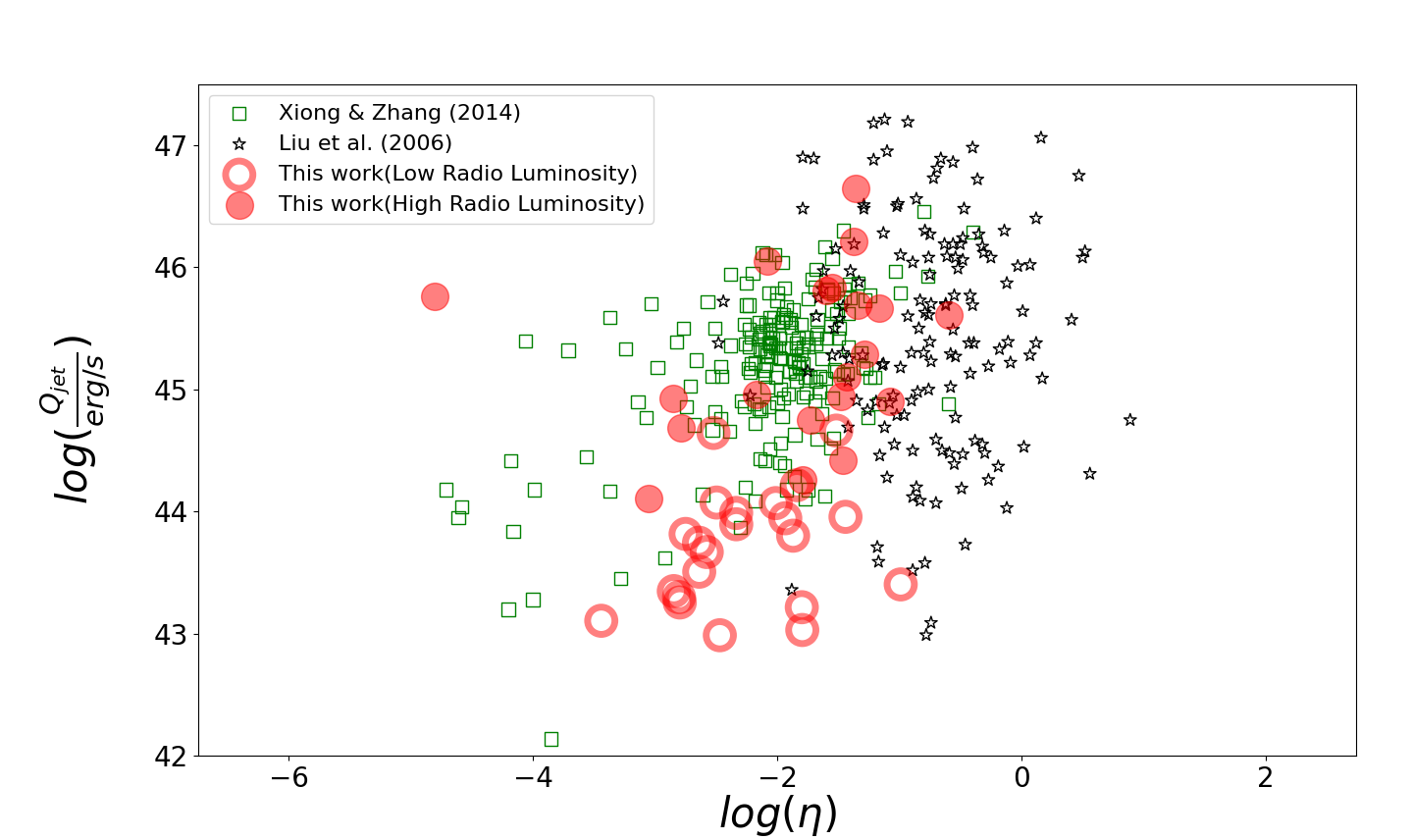}
        \caption{Plot of the jet power vs. the Eddington ratio ($\eta$)  of the associated black hole. In addition to our sample, the data found in \cite{xiong_intrinsic_2014} and \cite{liu_jet_2006} are plotted.}
        \label{fig:Qjet-eta}
\end{center}
\end{figure}

Figure~\ref{fig:Qjet-eta} displays instead the jet power, plotted against the Eddington ratio $\eta$ (see Sect.~\ref{sec:Edd_ratio}) for our radio sources and the galaxies in the comparison sample \citep{xiong_intrinsic_2014,liu_jet_2006}. Higher accretion rates favor more powerful jets to be launched by the central engine, as indeed the jet power increases with increasing Eddington ratio. For our sample of radio sources, we find that the two quantities are well correlated at a level of 2.9-$\sigma$ (${\rm p-value}=3.30\times10^{-3}$) by means of the Spearman test. No clear distinction in terms of $\eta$ is found between the two classes of low- and high-luminosity radio sources, which are distinguished in Fig.~\ref{fig:Qjet-eta}. 
However, as pointed out in Sect.~\ref{sec:Edd_ratio} our radio sources have an Eddington ratio of $\log \eta = -1.9$ on average. This value is typical of radiatively efficient accretion disks, such as the \citet{Shakura_Sunyaev1973} optically thick and geometrically thin accretion disk, which is commonly invoked to explain the optical-UV emission in type~I AGN \citep[e.g.,][]{Ghisellini2010,castignani_black-hole_2013}.

We can then estimate the accretion rate $\dot{M}= L_{\rm disk}/(\epsilon\;{\rm c}^2)$, where $\epsilon$ is the mass-to-light conversion efficiency, for which we adopted the standard value $\epsilon = 0.1$, which is typical of radiatively efficient disks. For our radio sources, we obtain a median (mean) accretion rate of 0.16~$M_\odot$~yr$^{-1}$ (0.6~$M_\odot$~yr$^{-1}$), which corresponds to a substantial SMBH mass growth of $\Delta M_{\rm BH} = 1.6\times10^6~M_\odot$ ($6.0\times10^6~M_\odot$) over an AGN duty cycle with typical duration of $\sim10^7$~yr. 

Altogether, the fact that the SMBHs of the radio sources in our sample accrete at a sub-Edddington rate, regardless of their redshift, suggests that most of their mass has likely been built up at earlier epochs. 
Furthermore, while on one hand, the observed accretion state sustains  both the nuclear activity and the SMBH growth at subparsec scales, on the other hand, it also ultimately favors the persistence of large-scale radio jets, which may prevent the host galaxy from accreting gas at kiloparsec scales and thus form stars effectively. This radio-mode AGN feedback may be responsible for the presence of overmassive SMBHs in our sample of radio-loud AGN. It is worth mentioning that  the five $z\sim0.37-0.95$ radio sources with high $M_{\rm BH}/M_\star$ ratios that we discussed in Sect.~\ref{sec:BH_vs_Mstar} accrete a sub-Eddington rate of $\eta\sim1\%$, while they have a normal jet power $Q_{\rm jet}\sim10^{44}$~erg~s$^{-1}$ on average.

\subsection{Radio-loud AGN and their environments}
A substantial fraction $(17\pm5)\%$ of our radio sources have high $M_{\rm BH}/M_\star$ ratios (Sect.~\ref{sec:BH_vs_Mstar}) and may be early-type galaxies at the center of clusters. For these galaxies, the stellar mass assembly may have been  halted, with reduced star formation activity, as typically found in cluster core ellipticals, while their black holes continue to grow via accretion. This interpretation is consistent with earlier studies \citep[e.g.,][]{Trujillo2014} as well as with the substantial fraction (19\%) of radio sources in our sample with low ${\rm SFR}<5~M_\odot$/yr, while many others have SFR upper limits.  It is indeed known that cluster-core early-type galaxies tend to be outliers in the  $M_{\rm BH}$ vs $M_{\rm bulge}$ relation \citep[e.g.,][]{mcconnell_revisiting_2013}. 
A famous example of a possibly overmassive black hole is the case of NGC~1277, which is a lenticular galaxy in the Perseus cluster \citep[e.g.,][]{van_den_Bosch2012,Emsellem2013,Scharwachter2016}.

{Motivated by these studies, we searched in the literature for clusters around the radio sources in our sample, searching by coordinates in the NASA/IPAC Extragalactic Database (NED). NED includes several catalogs of clusters that were identified in wide-field surveys \citep{Goto2002,Koester2007,McConnachie2009,Hao2010,Durret2011,Durret2015,Wen2012,Radovich2017,Rykoff2012,Oguri2018}. Our search yielded three matches. Radio sources RS~49, G~372455, and G~748815 are in the cores (at cluster-centric distances $\lesssim0.5$~Mpc) of the clusters [LIK2015]~J034.16359-04.73395 \citep[$z=0.89$;][]{Lee2015}, WHL~J090325.6+011215 \citep[$z=0.31$;][]{Wen2012,Wen2015}, and HSCS~J142538+002320 \citep[$z=0.33$;][]{Oguri2018}, respectively. These clusters have redshifts that are consistent with those of the radio sources as well as richness-based masses $M_{200}\sim(0.9-3.0)\times10^{14}~M_\odot$. These are therefore moderately massive clusters at intermediate to high redshifts. 
The three associated radio sources instead have moderately overmassive black holes $\log(M_{\rm BH}/M_\odot)\simeq8.6-9.3$, in particular, in comparison to the stellar masses of the systems  $-2.65\lesssim\log(M_{\rm BH}/M_\star)\lesssim-2.24$ (see Fig.~\ref{fig:MBH-M*-z}). 

These results support the interpretation described above that the cluster environments tend to prevent the stellar mass assembly of cluster early-type galaxies, resulting in observed overmassive black holes. Nevertheless, it is worth mentioning  that only three radio sources of our sample are found in clusters. However, this is expected as clusters at higher redshifts ($z\gtrsim1$) or with lower masses $M_{200}\lesssim1\times10^{14}~M_\odot$ typical of rich groups are more difficult to find with current surveys and observational facilities. It is therefore possible that additional galaxies are hosted in clusters, as distant radio sources are often found in dense megaparsec-scale environments \citep[e.g.,][]{Galametz2012,Castignani2014,Malavasi2015,Golden-Marx2019,Moravec2020}.}

\section{Discussion and conclusions}
\label{sec:conclusions}
We have  investigated the evolution of distant radio-loud AGN, as well their coevolution with their host galaxies and the SMBHs at their center. To do this, we built a sample of 42 radio-loud AGN with spectroscopic redshifts between $z\sim0.3-3.8$ by cross-matching the 1.4~GHz VLA FIRST  point-source catalog with available IR to optical spectrophotometric surveys including SDSS and DES in the optical, WISE in the IR, and the Galaxy And Mass Assembly (GAMA) spectroscopic survey.
As we are interested in assessing the SMBH masses, the 42 galaxies were further selected by requiring broad emission lines in \halpha, \hbeta, \mgii, or \civ, with an ${\rm FWHM}>1000$~km~s$^{-1}$. Based on the available multiwavelength photometry, we modeled the SEDs of the sources in the sample, and then derived estimates of the stellar mass ($M_\star$) and SFR for all sources, while for GAMA sources, we took them from the literature. 
We find that the 42 radio sources are broadly consistent with the star-forming main sequence.

For all sources, we then  estimated i) the black hole mass $M_{\rm BH}$, based on single-epoch broad-line region spectra, ii) the ratio of black hole to stellar mass, $M_{\rm BH}/M_\star$, iii) the jet power $Q_{\rm jet}$, on the basis of the low-frequency radio continuum emission, and iv) the Eddington ratio $\eta$. 
Although samples of distant AGN with SMBH mass estimates are rapidly growing \citep[e.g.,][]{shen_catalog_2011,shaw_spectroscopy_2012,Dabhade2020,Rakshit2020,Gloudemans2021,Li2022}, our study still represents one of the first in which all these quantities are derived simultaneously for a single sample of distant radio-loud AGN.

Our radio sources have $\log(M_{\rm BH}/M_\odot)\simeq7.1-10.0$ and high $\log(M_\star/M_\odot)\simeq10.2-12.0$, which agrees with the fact that radio-loud AGN are almost invariably associated with the most massive galaxies and black holes \citep[e.g.,][]{best_host_2005,Chiaberge_Marconi2011}.  While overall our sources follow the expected trends previously found in the literature, a substantial fraction of our sources, 9 out of 42 (i.e., 21\%), have black hole masses $\log(M_{\rm BH}/M_\odot)>9$ well above the values predicted by the scaling relations \citep{haring_black_2004,sani_spitzerirac_2011,degraf_scaling_2015}. In particular, five sources out of the nine (12$\%$ of the full radio source sample) are clearly  overmassive outliers, with $M_{\rm BH}/M_\star > 2\%$. This fraction is remarkably higher than that of $1.6\%$ found for AGN at similar redshifts from the literature. These overmassive SMBHs  are thus the high-$z$ counterparts of overmassive SMBHs found in previous studies of nearby early-type galaxies \citep[e.g.,][]{mcconnell_revisiting_2013, Trujillo2014}.

Our results imply that the growth of black hole masses in at least a substantial fraction of radio-loud AGN largely occurs at early epochs, while the early stellar mass assembly may not be so efficient. This population of radio-loud AGN with high $M_{\rm BH}/M_\star$ ratios have likely experienced a different stellar mass growth than other types of AGN, and we further investigated this scenario in terms of additional complementary probes.

Following early studies on nearby galaxies \citep[e.g.,][]{mcconnell_revisiting_2013, Trujillo2014}, we found that three of our radio-loud AGN with moderately overmassive SMBHs are hosted in clusters from the literature, while clusters and groups around the majority of the remaining radio-loud AGN will likely be detected with forthcoming surveys such as {\it Euclid} \citep{Adam2019}.  These results suggest that the cluster environments tend to prevent the stellar mass assembly of cluster early-type galaxies, possibly via radio-mode AGN feedback.

We found that the nuclear accretion and jet properties of the SMBHs of the radio sources in our sample accrete at a sub-Eddington rate ($\eta\sim1\%$) on average, where higher accretion rates favor more powerful jets to be launched by the central engine. We also find that high jet powers ($Q_{\rm jet} \gtrsim10^{45}$~erg~s$^{-1}$) are invariably associated with high radio luminosity sources ($L_{\rm 1.4~GHz}>2\times10^{32}$~erg~s$^{-1}$~Hz$^{-1}$). Altogether, the observed accretion state sustains both the nuclear activity and the SMBH growth at subparsec scales, while it ultimately favors the persistence of large-scale radio jets, which may prevent the host galaxy from accreting gas at kiloparsec scales and thus form stars effectively. Radio-mode AGN feedback may be responsible for the presence of overmassive SMBHs in our sample of radio-loud AGN.

Targeted observations of the ionized and the molecular gas are nevertheless needed to further investigate the proposed radio-mode AGN feedback scenario. Future studies on larger and higher-redshift samples of radio-loud AGN will become possible with the advent of forthcoming radio to optical surveys
such as those with the Vera Rubin Observatory and {\it Euclid} in the IR-optical,  SKA in radio, and its pathfinders and precursors (LOFAR, ASKAP, and MeerKAT).

\begin{acknowledgements}
We thank the anonymous referee for helpful comments which contributed to improve the paper.
GC  acknowledges the support from the grants ASI n.2018-23-HH.0 and  PRIN-MIUR 2017 WSCC32.
We thank Christophe Benoist for helpful discussion about the exploitation of DES data.
%GAMA, SDSS: it is enough to cite the papers
%WISE
This publication makes use of data products from the Wide-field Infrared Survey Explorer, which is a joint project of the University of California, Los Angeles, and the Jet Propulsion Laboratory/California Institute of Technology, and NEOWISE, which is a project of the Jet Propulsion Laboratory/California Institute of Technology. WISE and NEOWISE are funded by the National Aeronautics and Space Administration.
%NED
This research has made use of the NASA/IPAC Extragalactic Database (NED),
which is operated by the Jet Propulsion Laboratory, California Institute of Technology,
under contract with the National Aeronautics and Space Administration.
%DES
This project used public archival data from the Dark Energy Survey (DES). Funding for the DES Projects has been provided by the U.S. Department of Energy, the U.S. National Science Foundation, the Ministry of Science and Education of Spain, the Science and Technology FacilitiesCouncil of the United Kingdom, the Higher Education Funding Council for England, the National Center for Supercomputing Applications at the University of Illinois at Urbana-Champaign, the Kavli Institute of Cosmological Physics at the University of Chicago, the Center for Cosmology and Astro-Particle Physics at the Ohio State University, the Mitchell Institute for Fundamental Physics and Astronomy at Texas A\&M University, Financiadora de Estudos e Projetos, Funda{\c c}{\~a}o Carlos Chagas Filho de Amparo {\`a} Pesquisa do Estado do Rio de Janeiro, Conselho Nacional de Desenvolvimento Cient{\'i}fico e Tecnol{\'o}gico and the Minist{\'e}rio da Ci{\^e}ncia, Tecnologia e Inova{\c c}{\~a}o, the Deutsche Forschungsgemeinschaft, and the Collaborating Institutions in the Dark Energy Survey.
The Collaborating Institutions are Argonne National Laboratory, the University of California at Santa Cruz, the University of Cambridge, Centro de Investigaciones Energ{\'e}ticas, Medioambientales y Tecnol{\'o}gicas-Madrid, the University of Chicago, University College London, the DES-Brazil Consortium, the University of Edinburgh, the Eidgen{\"o}ssische Technische Hochschule (ETH) Z{\"u}rich,  Fermi National Accelerator Laboratory, the University of Illinois at Urbana-Champaign, the Institut de Ci{\`e}ncies de l'Espai (IEEC/CSIC), the Institut de F{\'i}sica d'Altes Energies, Lawrence Berkeley National Laboratory, the Ludwig-Maximilians Universit{\"a}t M{\"u}nchen and the associated Excellence Cluster Universe, the University of Michigan, the National Optical Astronomy Observatory, the University of Nottingham, The Ohio State University, the OzDES Membership Consortium, the University of Pennsylvania, the University of Portsmouth, SLAC National Accelerator Laboratory, Stanford University, the University of Sussex, and Texas A\&M University.
Based in part on observations at Cerro Tololo Inter-American Observatory, National Optical Astronomy Observatory, which is operated by the Association of Universities for Research in Astronomy (AURA) under a cooperative agreement with the National Science Foundation.
\end{acknowledgements}

\bibliographystyle{aa}
\bibliography{RLAGN2}

\begin{appendix}
\section{Tables}
We report below some tables summarizing the properties of the radio sources in our sample.

\renewcommand{\arraystretch}{2}
\begin{table*}[h]
        \vspace{-0cm}
        \hspace*{-0cm}
        \begin{adjustbox}{width=1.0\textwidth, left}
        \label{tab:sample}
        \begin{tabular}{ccccccccccc}
                \hline \hline
                 ID & RA & Dec. & $z$ & $\log\frac{L_{1.4~\rm GHz}}{{\rm erg~s}^{-1}~{\rm Hz}^{-1}}$ & $\log(L_{\rm dust}/L_\odot)$ & $\log(M_{\star}/M_\odot)$  & SFR &
                 WISE class & Type & Name       \\
                  & hh:mm:ss.ss & dd:mm:ss.ss & &  &  & & ($M_\odot$/yr) &  &  \\
                                  \hline
                  (1) & (2) & (3) & (4) & (5) &  (6) & (7) & (8) & (9) & (10) & (11)              \\
                  \hline
                  RS 16 & 02:23:34.94 & -06:37:22.99 & 1.217 & $33.65\pm0.01$ & 11.48 & 10.40 & 32 & AGN & Radio Source & SDSS J022334.90-063722.9\\ 
                  RS 49 & 02:16:40.74 & -04:44:04.97 & 0.875 & $33.51\pm0.01$ & 13.65 & 11.95 & $<$4801 & AGN & QSO & FBQS~J0216-0444 \\ 
                  RS 52 & 02:19:38.89 & -05:18:05.28 & 2.207 &  $32.65\pm0.04$ & 12.34 & 11.73 & $<$235 & AGN & QSO & 3XMM~J021938.8-051805 \\
                  RS 60 & 02:19:54.63 & -05:49:22.32 & 0.322 & $30.63 \pm 0.05$ & 9.08 & 10.15 & 0.13 & AGN & QSO & SDSS J021954.62-054922.2 \\
                  RS 62 & 02:22:47.91 & -04:33:31.02 & 1.635 & $32.60\pm0.02$ & 12.43 & 11.02 & $<$289 &  AGN & QSO & XXL-N 027$\_$020 \\
                  RS 79 & 02:22:55.95 & -05:18:15.85 & 1.756 & $34.63\pm0.01$  & 12.32 & 11.48 &  $<$225 & AGN & QSO  & PKS~0220-055 \\ 
                  RS 81 & 02:26:07.42 & -05:32:09.52 & 0.779 & $32.30\pm0.01$ & 11.53 & 10.76 & 36 &  Intermediate disk & X-ray source &      XXL-N 062$\_$013 \\ 
                  RS 82 & 02:25:56.39 & -05:34:51.44 & 2.879 & $33.68\pm0.01$ & 12.75 & 11.25 & $<$605 & AGN & QSO & XXL-N 062$\_$009 \\ 
                  RS 83 & 02:25:05.12 & -05:36:47.94 & 0.682 & $33.60\pm0.01$ & 9.58 & 11.11 & 0.41 & Intermediate disk & QSO & PMN~J0225-0536 \\
                  RS 104 & 02:27:12.99 & -04:46:36.36 & 0.981 & $32.32\pm0.01$ & 12.62 & 11.15 & $<$448 & AGN & X-ray source & SPIRE~13430 \\ 
                  RS 113 & 02:45:31.51 &  -00:26:12.32 & 2.082 &  $32.83\pm0.02$ & 12.52 & 11.75 & $<$356 & AGN & QSO & 2SLAQ~J024531.53-002612.2 \\
                  RS 151 & 02:27:40.56 & -04:02:51.16 & 2.603 & $32.42\pm0.07$ & 13.69 & 11.54 & $<$5265 & AGN & QSO & XXL-N 044$\_$070\\ 
                  RS 159 & 02:29:15.79 & -04:42:15.95 & 1.074 & $34.12\pm0.01$ & 12.29 & 10.28 & <21 & AGN & QSO & 3XMM J022915.7-044216 \\
                  RS 177 & 02:51:56.32 & +00:57:06.53 & 0.471 & $31.80\pm0.01$ & 11.74 & 10.45 & $<$59 & AGN & QSO & LBQS~0249+0044\\ 
                  RS 190 & 02:51:15.50 & +00:31:35.45 & 1.978 & $33.93\pm0.01$ & 12.02 & 11.10 & <114 & AGN & QSO & WISEA J025115.50+003135.4 \\
                  RS 195 & 02:47:06.66 & +00:23:18.10 & 0.363 & $30.85\pm0.03$ & 11.67 & 10.94 & $<$50 & AGN & QSO & FBQS~J0247+0023 \\ 
                  RS 197 &  02:46:16.61 & +00:19:53.11 & 3.791 & $33.46\pm0.02$ & 11.61 & 11.11 & 44 & Starburst/LIRG & QSO & SDSS~J024616.60+001953.6 \\ 
                  RS 205 & 02:53:40.94 & +00:11:10.04 & 1.683 & $33.04\pm0.01$ & 12.73  & 11.23 &  $<$577 & AGN & QSO & LBQS~0251-0001\\ 
                  RS 206 & 02:48:54.80 & +00:10:53.84 & 1.145 &  $33.49\pm0.01$ & 12.92 & 11.05  & $<$904 & AGN & QSO & 2SLAQ~J024854.80+001053.9 \\ 
                  RS 214 & 02:50:48.66 & +00:02:07.46  & 0.766 & $32.67\pm0.01$ & 11.93 & 10.43 &  $<$91 & AGN & QSO & FBQS~J0250+0002 \\ 
                  RS 237 & 02:49:23.22 & -00:54:38.04 & 0.953 &  $31.63\pm0.06$ & 11.69 & 10.27 & $<$53 & AGN & QSO & 2SLAQ~J024923.20-005437.7 \\ 
                
                  G 55673  & 12:11:55.32 & -00:20:19.38 & 0.436 &  $33.16\pm0.04$ & 11.55 & 11.19 & 23 & Starburst/LIRG & G & SDSS J121155.31-002019.4 \\
          G 71277  & 12:16:12.27 &  00:04:17.91 & 0.316 & $32.29\pm0.03$  & 10.01 & 11.23 & 0.49 & Intermediate disk & G & SDSS J121612.26+000417.8 \\ 
          G 165213 & 12:01:13.77 & -02:42:41.34 & 0.307 &  $31.90\pm0.08$  & 11.64 & 11.27 & 34 & AGN & QSO & SDSS J120113.76-024241.3\\
          G 196970 & 08:53:52.21 & -00:45:31.12 & 0.323 & $31.99\pm0.06$  & 10.00 & 11.08 & 0.74 & Intermediate disk & G & SDSS J085352.21-004531.1\\
          G 208794 & 08:40:44.47 &  00:03:05.27 & 0.449 & $31.84\pm0.10$  & 11,59 & 10.92 & 28 & AGN & G & SDSS J084044.10+000307.2\\
          G 249591 & 14:08:22.42 &  02:08:53.70 & 0.432 & $32.45\pm0.02$  & 10.64 & 11.31 & 4.3 & Intermediate disk & G & SDSS J140822.76+020853.1\\
          G 251343 & 14:32:57.38 &  02:03:24.46 & 0.761 & $32.66\pm0.01$  & 12.22 & 11.54 & $<$1119 & Starburst/LIRG & G & SDSS J143257.64+020321.3\\
          G 298359 & 14:37:31.86 &  01:18:58.14 & 0.342 & $34.06\pm0.01$  & 10.81 & 10.92 & 3.1 & Intermediate disk & Radio Source & GAMA J143257.41+020329.1\\
          G 372455 & 09:03:25.57 &  01:12:14.86 & 0.311 & $32.68\pm0.01$  & 9.842 & 11.02 & 0.75 & Intermediate disk & Radio Source & GAMA J090325.48+011214.1\\
          G 537618 & 12:23:48.39 & -00:52:50.43 & 0.490 & $32.57\pm0.02$  & 12.06 & 10.79 & 43 & Starburst/LIRG & Radio Source & SDSS J122347.89-005249.2\\
          G 714133 & 14:25:33.03 &  01:07:37.73 & 0.556 & $32.48\pm0.02$  & 12.35 & 10.80 & $<$238 & AGN & Radio Source & NVSS J142533+010739\\
          G 714228 & 14:31:20.49 &  01:14:56.44 & 0.343 &  $33.10\pm0.01$  & 10.33 & 10.67 & 5.9 & AGN & G & SDSS J143120.07+011459.2\\
          G 720847 & 08:47:02.80 &  01:30:01.50 & 0.417 & $32.87\pm0.01$ & 10.45 & 11.11 & 30 & AGN & QSO & WISEA J084702.78+013001.5\\
          G 721940 & 14:21:30.03 &  02:13:02.43 & 0.640 & $32.67\pm0.01$  & 11.89 & 10.94 & $<$75 & Intermediate disk & G & SDSS J142130.60+021308.9\\
          G 745066 & 14:51:22.48 & -00:33:41.05 & 0.377 & $32.06\pm0.06$  & 11.81 & 11.72 & 45 & AGN & QSO & WISEA J145122.47-003341.0\\
          G 746605 & 12:21:02.95 & -00:07:33.74 & 0.364 &  $31.78\pm0.10$  & 12.08 & 10.70 & $<$131 & AGN & G & SDSS J122103.51-000749.1\\
          G 748144 & 11:39:54.20 &  00:13:47.26 & 0.589 & $32.37\pm0.03$  & 12.09 & 11.67 & $<$109 & AGN & Radio Source & GAMA J113952.95+001348.6\\
          G 748815 & 14:25:45.91 &  00:22:42.73 & 0.326 & $33.82\pm0.01$  & 11.31 & 10.87 & 14 & AGN & QSO & WISEA J142545.90+002242.7\\
          G 804203 & 09:20:53.32 &  00:03:53.94 & 0.506 &  $32.37\pm0.03$  & 11.93 & 10.79 & 41 & Starburst/LIRG & G & SDSS J092053.32+000353.9\\
          G 835899 & 08:42:16.99 &  01:09:17.87 & 0.762 & $32.05\pm0.06$  & 12.19 & 11.83 & $<$70 & AGN & G & WISEA J084217.25+010834.6\\
          G 887308 & 14:37:01.00 & -01:03:49.03 & 0.547 &  $32.28\pm0.03$  & 10.47 & 11.55 & 2.4 & Intermediate disk & G & SDSS J143702.15-010357.2\\
                \hline
                \end{tabular}
    \end{adjustbox}
    \caption{Properties of the radio sources in our sample. Column (1): Galaxy ID; (2-3) RA and Dec. coordinates; (4) Spectroscopic redshift; (5) 1.4~GHz rest-frame luminosity density; (6-7) SED-based dust luminosity and stellar mass; (8) SFR; (9) WISE color-based class; (10-11) source type and name as found in the NED.}
    \label{tab:A1sample}
\end{table*}

\renewcommand{\arraystretch}{2}
\begin{table*}[h]
        \vspace{-0cm}
        \hspace*{-0cm}
        \begin{adjustbox}{width=1.0\textwidth, left}
                \label{tab:prop}
        \scriptsize
        \begin{tabular}{ccccccccccc}
                                \hline \hline
                 ID & $z$ & Line & FWHM & $\log(L_{\rm line}/({\rm erg~s}^{-1}))$ & $\log(M_{\rm BH}/M_\odot)$ & $\log(Q_{\rm jet}/({\rm erg~s}^{-1}))$ & $\log(L_{\rm BLR}/({\rm erg~s}^{-1}))$ & $\log~\eta$ \\
                  & & & ($10^3$~km~s$^{-1}$) & \\
                  \hline
                          (1) & (2) & (3) & (4) & (5) &  (6) & (7) & (8) & (9)            \\
                                \hline
             RS 16    & 1.22 & \mgii   & 4.08 $\pm$ 0.42 & 42.53 $\pm$ 0.07 & 8.00 $\pm$ 0.12 & 45.81 & 43.50 & -1.60\\
                    RS 49    & 0.88 & \hbeta & 6.77 $\pm$ 0.11 & 44.02 $\pm$ 0.01 & 9.30$\pm$ 0.04 & 45.69 & 45.06 & -1.34\\
                    RS 52    & 2.21 & \mgii   & 6.00 $\pm$ 0.43 & 43.41 $\pm$ 0.14 & 8.88 $\pm$ 0.13 & 44.94 & 44.51 & -1.48\\
                    RS 60    & 0.32 & \halpha& 1.74 $\pm$ 0.07  & 41.88 $\pm$ 0.02 & 7.13 $\pm$ 0.04 & 43.22 & 42.44 & -1.80 \\
                    RS 62    & 1.64 & \mgii   & 3.20 $\pm$ 0.11 & 43.26 $\pm$ 0.02 & 8.25 $\pm$ 0.08 & 44.90 & 44.27  & -1.07\\
                    RS 79    & 1.76 & \mgii   & 6.48 $\pm$ 0.16 & 43.47 $\pm$ 0.03 & 8.99 $\pm$ 0.08 & 46.60 & 44.73 & -1.30\\
                    RS 81    & 0.78 & \mgii   & 12.47 $\pm$ 2.07& 42.14 $\pm$ 0.14 & 8.72 $\pm$ 0.18 & 44.60 & 43.30  & -2.52\\
                    RS 82    & 2.88 & \civ    & 5.26  $\pm$ 0.37 & 43.49 $\pm$ 0.12 & 8.73 $\pm$ 0.24 & 45.83 & 44.28 & -1.55\\
                    RS 83    & 0.68 & \hbeta & 7.03  $\pm$ 1.67 & 41.87 $\pm$ 0.14 & 8.28 $\pm$ 0.23 & 45.70 & 40.57 & -4.80\\
                    RS 104   & 0.98 & \hbeta & 6.31  $\pm$ 0.30 & 43.47 $\pm$ 0.03 & 8.97 $\pm$ 0.06 & 44.60 & 44.56  & -1.52\\
                    RS 113   & 2.08 & \mgii   & 5.89  $\pm$ 0.24 & 43.62 $\pm$ 0.06 & 9.00 $\pm$ 0.09 & 45.10 & 44.67 & -1.43\\
                    RS 151   & 2.60 & \mgii   & 7.68  $\pm$ 0.30 & 44.47 $\pm$ 0.08 & 9.76 $\pm$ 0.09 & 44.75 & 45.13  & -1.73\\
                    RS 159   & 1.07 & \mgii   & 3.95  $\pm$ 0.21 & 42.94 $\pm$ 0.03 & 8.22 $\pm$ 0.10 & 46.21 & 43.95 & -1.37 \\
                    RS 177   & 0.47 & \hbeta & 5.31  $\pm$ 0.14  & 42.60 $\pm$ 0.01 & 8.39 $\pm$ 0.06 & 44.22 & 43.65  & -1.84\\
                    RS 190   & 1.98 & \civ    & 14.66 $\pm$ 1.29 & 44.40 $\pm$ 0.10 & 10.03 $\pm$ 0.24 & 46.05 & 45.05 & -2.08\\
                    RS 195   & 0.36 & \hbeta & 2.90  $\pm$ 0.03   & 42.42 $\pm$ 0.01 & 7.78 $\pm$ 0.06 & 43.40 & 43.89 & -0.99\\
                    RS 197   & 3.79 & \civ    & 1.93  $\pm$ 0.39 & 43.64 $\pm$ 0.09 & 7.92 $\pm$ 0.28 & 45.61 & 44.43 & -0.50\\
                    RS 205   & 1.68 & \mgii   & 7.22  $\pm$ 0.18 & 43.93 $\pm$ 0.02 & 9.38 $\pm$ 0.07 & 45.28 & 45.19 & -1.29\\
                    RS 206   & 1.15 & \mgii   & 4.12  $\pm$ 0.11 & 43.54 $\pm$ 0.01 & 8.64 $\pm$ 0.07 & 45.60 & 44.58 & -1.10\\
                    RS 214   & 0.77 & \hbeta & 11.05 $\pm$ 0.48 & 43.42 $\pm$ 0.02 & 9.43 $\pm$ 0.06 & 44.96 & 44.36 & -2.10\\
                    RS 237   & 0.95 & \mgii   & 9.89  $\pm$ 0.99 & 43.24 $\pm$ 0.07& 9.21 $\pm$ 0.12 & 44.07 & 44.30 & -2.01\\
            G 55673  & 0.44 & \hbeta & 1.20  $\pm$ 0.19   & 41.66 $ \pm $ 0.09 & 8.45 $ \pm $ 0.16 & 43.51 & 33.91 & -2.60\\
            G 71277  & 0.32 & \halpha& 1.05  $\pm$ 0.11   & 41.99 $ \pm $ 0.05 & 8.36 $ \pm $ 0.09 & 43.35 & 33.61 & -2.85\\
            G 165213 & 0.31 & \halpha& 1.19  $\pm$0.93  & 42.79 $ \pm $ 0.01 & 8.82 $ \pm $ 0.71 & 42.99 & 34.45 & -2.47\\
            G 196970 & 0.32 & \hbeta & 1.20  $\pm$0.60$^{a}$   & 40.89 $ \pm $ 0.07 & 8.08 $ \pm $ 0.45 & 43.11 & 32.74 & -3.44\\
            G 208794 & 0.45 & \hbeta & 1.20  $\pm$0.11   & 41.35 $ \pm $ 0.04 & 8.30 $ \pm $ 0.12 & 43.26 & 33.6 & -2.80\\
            G 249591 & 0.43 & \hbeta & 1.20  $\pm$0.62$~^{a}$   & 41.66 $ \pm $ 0.14 & 8.45 $ \pm $ 0.46 & 43.75 & 33.91 & -2.64\\
            G 251343 & 0.76 & \hbeta & 1.19  $\pm$0.15$^{a}$   & 43.96 $ \pm $ 0.02 & 9.57 $ \pm $ 0.12 & 44.42 & 36.21 & -1.46\\
            G 298359 & 0.34 & \halpha& 1.19  $\pm$0.18  & 42.07 $ \pm $ 0.06 & 8.51 $ \pm $ 0.14 & 44.93 & 33.76 & -2.85\\
            G 372455 & 0.31 & \halpha& 1.19  $\pm$0.60$^{a}$   & 42.66 $ \pm $ 0.08 & 8.77 $ \pm $ 0.08 & 43.67 & 34.29 & -2.58\\
            G 537618 & 0.49 & \hbeta & 1.10  $\pm$ 0.04   & 43.86 $ \pm $ 0.02 & 9.45 $ \pm $ 0.05 & 43.96 & 36.11 & -1.44\\
            \hline
            \end{tabular}
    \end{adjustbox}
\end{table*}
 \renewcommand{\arraystretch}{2}
\begin{table*}[h]
        \vspace{-0cm}
        \hspace*{-0cm}
        \begin{adjustbox}{width=1.0\textwidth, left}
        \scriptsize
        \begin{tabular}{ccccccccccc}
                \hline \hline
                 ID & $z$ & Line & FWHM & $\log(L_{\rm line}/({\rm erg~s}^{-1}))$ & $\log(M_{\rm BH}/M_\odot)$ & $\log(Q_{\rm jet}/({\rm erg~s}^{-1}))$ & $\log(L_{\rm BLR}/({\rm erg~s}^{-1}))$ & $\log~\eta$ \\
                  & & & ($10^3$~km~s$^{-1}$) & \\
                  \hline
                          (1) & (2) & (3) & (4) & (5) &  (6) & (7) & (8) & (9)            \\
                                \hline
                        G 714133 & 0.56 & \hbeta & 1.20  $\pm$0.19  & 42.25 $ \pm $ 0.01 & 8.74 $ \pm $ 0.10 & 43.99 & 34.50 & -2.34\\
            G 714228 & 0.34 & \hbeta & 1.20  $\pm$0.11   & 40.86$\pm$0.06 & 8.06 $ \pm $ 0.13 & 44.11 & 33.11 & -3.05\\
            G 720847 & 0.42 & \hbeta & 1.19  $\pm$0.24  & 41.92 $ \pm $ 0.02 & 8.57 $ \pm $ 0.19 & 44.07 & 34.17 & -2.5\\
            G 721940 & 0.64 & \hbeta & 1.20  $\pm$0.60$^{a}$   & 43.31 $ \pm $ 0.04 & 9.25 $ \pm $ 0.44 &  44.26 & 35.56 & -1.79\\ 
            G 745066 & 0.38 & \hbeta & 1.19 $\pm$0.60$^{a}$ & 41.73 $ \pm $ 0.02 & 8.47 $ \pm $ 0.45 & 43.30 & 33.77 & -2.8\\ 
            G 746605 & 0.37 & \hbeta & 1.24  $\pm$ 0.09   & 43.37 $ \pm $ 0.04 & 9.31 $ \pm $ 0.08 & 43.03 & 35.62 & -1.79\\
            G 748144 & 0.59 & \hbeta & 1.22  $\pm$ 0.31  & 43.08$ \pm $ 0.09 & 9.16 $ \pm $ 0.23 & 43.95 & 35.33 & -1.93\\
            G 748815 & 0.33 & \hbeta & 1.19  $\pm$0.07$^{a}$   & 42.05$ \pm $0.01 & 8.63 $ \pm $ 0.09 & 44.68 & 33.94 & -2.79\\
            G 804203 & 0.51 & \hbeta & 1.20   $\pm$0.60$^{a}$   & 41.44$ \pm $0.11 & 8.34 $ \pm $ 0.45 & 43.82 & 33.69 & -2.75\\ 
            G 835899 & 0.76 & \hbeta & 1.19  $\pm$0.16$^{a}$  & 42.24$ \pm $0.02 & 8.72 $ \pm $ 0.13 & 43.90 & 34.49 & -2.33\\
            G 887308 & 0.55 & \hbeta & 1.20   $\pm$0.19$^{a}$   & 43.17$ \pm $0.03 & 9.18 $ \pm $ 0.15 & 43.80 & 35.42 & -1.87\\
                \hline
                \end{tabular}
    \end{adjustbox}
    \caption{
    Black hole, accretion, and jet properties of the radio sources in our sample.
    Column (1): Source ID; (2) spectroscopic redshift; (3-5) broad emission line, FWHM, and line luminosity; (6) single-epoch SMBH mass; (7) jet power; (8) BLR luminosity; and (9) Eddington ratio. Sources marked with the symbol $^a$ in column (4) have uncertain FWHM errors from GAMA, when the lines were fit with a single Gaussian component. To overcome this limitation, we considered the GAMA fits that include both narrow and broad components of the emission line. We then assumed an FWHM relative error equal to that resulting from the fit to the broad component of the line.} 
    \label{tab:A2MBH}
\end{table*}

\end{appendix}
\end{document}